\documentclass[draftcls, 12pt, onecolumn]{IEEEtran}
\usepackage{bbm}
\usepackage{}
\usepackage{mathrsfs}
\usepackage{amssymb}
\usepackage{bbding}
\usepackage{amsfonts}
\ifCLASSINFOpdf

\else

\fi

\usepackage{graphicx,cite,epsfig,amssymb,amsmath,multirow,extarrows}
\usepackage{clrscode}
\usepackage{algorithm}
\usepackage{algpseudocode}
\usepackage{algpascal}
\usepackage{xcolor}
\usepackage{bm}
\usepackage{hyperref}
\usepackage{subfigure}
\usepackage{mathtools}
\usepackage{diagbox}
\usepackage[square, comma, sort&compress, numbers]{natbib}
\begin{document}

\title{Spectral Efficiency Analysis of Cell-Free Massive MIMO Systems with \\Zero-Forcing Detector}

\author{
Pei Liu, \IEEEmembership{Member,~IEEE},
Kai Luo, \IEEEmembership{Member,~IEEE},\\
Da Chen,
and Tao Jiang, \IEEEmembership{Fellow,~IEEE}

\thanks{P. Liu is with the School of Information Engineering, Wuhan University of Technology, Wuhan 430070, China, and also with Wuhan National Laboratory for Optoelectronics, School
of Electronic Information and Communications, Huazhong University of Science and Technology, Wuhan 430074, China (e-mail: pei.liu@ieee.org; peil@hust.edu.cn).}

\thanks{K. Luo, D. Chen, and T. Jiang  are with Wuhan National Laboratory for Optoelectronics, School
of Electronic Information and Communications, Huazhong University of Science and Technology, Wuhan 430074, China (e-mail: kluo@hust.edu.cn; chenda@hust.edu.cn; Tao.Jiang@ieee.org).}

}
\markboth{}
{}

\maketitle
\vspace{-8mm}
\begin{abstract}
In this paper,
we firstly derive two approximations of the achievable uplink rate with the perfect/imperfect channel state information (CSI) in cell-free massive multi-input multi-output (MIMO) systems, and all these approximations are not only in the simple, but also converge into the classical bounds achieved in conventional massive MIMO systems where the base-station (BS) antennas are co-located.
It is worth noting that the obtained two approximations with perfect CSI could be regarded as the special cases of the obtained two approximations with imperfect CSI when the pilot sequence power becomes infinite, respectively. Moreover, the theory analysis shows that all obtained approximations with perfect/imperfect CSI have an asymptotic lower bound $\frac{\alpha}{2}\log_2 L$ thanks to the extra {\emph {distance diversity}} offered by massively distributed antennas, where $L$ is the number of BS antennas and the path-loss factor $\alpha>2$, except for the free space environment. Obviously, these results indicate that the cell-free massive MIMO system has huge potential of spectral efficiency than the conventional massive MIMO system with the asymptotically tight bound $\log_2 L$.
\end{abstract}



\IEEEpeerreviewmaketitle

\section{Introduction}
\IEEEPARstart{T}{o} meet the ever-increasing demand for high data rate in future wireless communications, there is an urgent need to improve the spectral efficiency (SE) \cite{Qualcomm}. In this regard, massive multiple-input multiple-output (MIMO) technology has been at the forefront thanks to its high SE characteristic  provided by the massive array \cite{Marzetta10,Marzetta16,Hoydis13,Liu17,Zhang18}.  Conventional massive MIMO researches routinely assume all base-station (BS) antennas are placed at a fixed location \cite{Qiu18,Feng17}.  Recently, some cell-free massive MIMO systems (also referred to as distributed massive MIMO systems and large-scale distributed antenna systems) \cite{Ngo17,Li18,Guo162}, where massive BS antennas are distributed over a wide area to serve a amount of users, have been obtained a great deal of interest, due that they reduce the average distance between the user and BS antennas, resulting in additional features of improving coverage, saving power, offering more spatial resources, etc \cite{Dai11,Dai14}.



Obviously, the statistical characterization of the instantaneous channel correlation matrix is of vital importance to the SE performance based on the classic work performed by Telatar \cite{Telatar99}. For conventional massive MIMO systems, the centralized BS array is adopted that the large-scale fading coefficients are assumed to be identical, resulting in the instantaneous channel correlation matrix follows the Wishart distribution, and corresponding statistical properties and SE analysis have been well investigated in \cite{Liu17,Ngo13,Tulino04}.  However, in a cell-free massive MIMO system, the served user has different large-scale fading coefficients to all distributed BS antennas, resulting in its instantaneous channel correlation matrix  being modeled as a Gram matrix with each element having different variance. Recently, a toy example $2\times2$ Gram matrix was investigated in \cite{Auguin17}, where the probability density function (PDF) was obtained and its joint eigenvalues PDF was given in integral form. In summary, the SE analysis of cell-free massive MIMO systems has become exceedingly  more challenging and complex.


\subsection{Related Works}

On one hand, part of studies are performed with the non-orthogonal linear maximum-ratio combining/maximum-ratio transmission (MRC/MRT) processing. A  portion of the open literature adopts the structure of that the BS antennas are assigned to different multiple-antenna access points (APs). The analytical expressions of the achievable uplink/downlink rate were obtained in \cite{Li14} with imperfect channel state information (CSI), then, it could be achieved via the applications of the large dimensional random matrix theory \cite{Bai09}  and the worst-case assumption \cite{Hassibi03}. A asymptotic signal-to-interference-plus-noise ratio (SINR) expression was obtained in \cite{Zuo17}. Some bounds of the achievable downlink rate were obtained in \cite{Li18,Li181}.

Another deployment strategy is to let all BS antennas being randomly located in a large area. Considering the downlink with perfect CSI, an exact achievable downlink rate expression was given in special function form and its asymptotic performance was analyzed with an upper bound in \cite{Wang15}.  Apart from that, two closed-form expressions were proposed for the achievable uplink/downlink rates with imperfect CSI, and a near-optimal power allocation scheme was designed to greatly promote the uplink/downlink minimum rate in \cite{Ngo17}. A max-min SINR optimization problem was considered and the power allocation methods in the uplink and downlink were provided in \cite{Bashar18} and \cite{Nayebi17}, respectively.
Moreover,  the rate performance issue of downlink multicast was evaluated in \cite{Doan17}. It was implied that using downlink pilots can effectively improve the system performance \cite{Interdonato16}. A power control scheme was provided to reduce the power consumption in \cite{Ngo173}. Obviously, the SE analysis based on linear MRC/MRT processing has been largely and extensively characterized.

On the other hand,  some of studies pay attention on the orthogonal linear zero-forcing (ZF) processing. Considering a multiple-antenna user and distributed multiple-antenna APs with perfect CSI, an approximate closed-form expression of the achievable uplink rate was obtained in the Laguerre polynomial form based on the Gauss-Laguerre quadrature rule in \cite{Almelah17}. Moreover, an asymptotic closed-form achievable uplink rate expression that contains the Legendre function and Gauss hypergeometric function was provided in \cite{Yang15}. An asymptotic reliable rate was investigated  and a certain form of macro-multiplexing gain with its value interval for any user density was provided in \cite{Koyuncu18}. Moreover, by smartly utilizing the probability that two users are close to the same BS antenna is very low to approximate the effective channel gain, a lower bound of the achievable downlink rate was given in integral form and the corresponding scaling behavior was analyzed in \cite{Wang15}. With imperfect CSI, by wisely using the Gamma approximation, an upper bound of the ergodic achievable rate that is made up by Meijer's G-function was provided in \cite{Li18}. An effective SINR expression that is the function of small-scale fading, was given in \cite{Nayebi17}, and the similar case can be found in \cite{Nguyen17}. From the above, it becomes apparent that there are remarkably few results about the insightful analytical expression of the achievable uplink rate and the analysis about how system parameters affect the SE for both perfect CSI and imperfect CSI cases.


\subsection{Our Contributions}
Motivated by the previous discussion and benefited from the methodologies in previous works especially for downlink, in this paper, we mainly focus on the SE analysis of the ZF detector in cell-free massive MIMO systems with perfect CSI and imperfect CSI cases.

For the perfect CSI case, we study the achievable uplink rate performance and propose two approximate closed-form expressions of the achievable uplink rate which depend only on the large-scale fading coefficients and are in a reduced form.  In other words, these simple closed-form approximations make it possible to obtain insights about how system parameters affect the SE performance.
Particularly, it is important to note that the obtained two approximations can converge into the upper and lower bounds in conventional massive MIMO systems, respectively, when all BS antennas are at the same position. Further, by averaging out the large-scale fading coefficients, the asymptotic analysis reflects that the obtained two approximations of the achievable uplink rate have an asymptotic lower bound $\frac{\alpha}{2}\log_2 L$ when $L\rightarrow\infty$. Here, $L$ is the number of BS antennas and $\alpha$ denotes the path-loss factor that is often larger than 2, except for the free space environment. However, for the conventional massive MIMO systems, the  upper and lower bounds  have the asymptotically tight bound $\log_2 L$. It means that the cell-free massive MIMO systems have a better rate performance  than the conventional massive MIMO systems since cell-free massive MIMO offers extra {\emph {distance diversity}} thanks to the massively distributed antennas.

For the imperfect CSI case, the  similar results can be obtained as of perfect CSI case, but only with a little bit rate loss. Besides, it is found that the obtained two approximations in imperfect CSI case  converge into the two approximations in perfect CSI case when the pilot sequence power is large enough.
In particular, the main contributions of the paper are outlined as follows.
\begin{itemize}
    \item We investigate the achievable uplink rate performance for both perfect CSI and imperfect CSI based on ZF detector of the cell-free massive MIMO through deriving the new, tractable, and closed-form expressions of approximation of the achievable uplink rate.
    \item We observe that not only the imperfect CSI results can approach the perfect CSI results with high pilot sequence power, but also all approximations can reduce to the exact bounds in conventional massive MIMO when BS antennas are placed together.
    \item We analyze the asymptotic performance  and note that the obtained approximations have the asymptotic lower-bound $\frac{\alpha}{2}\log_2 L$ for both cases, which outperforms the conventional massive MIMO since its rate has the asymptotically tight bound $\log_2 L$.
\end{itemize}



\subsection{Organization and Notation}
The rest of the paper is organized as follows. The system model is described in Section II.
The analysis of SE with perfect CSI is presented in Section III.
The analysis of SE with imperfect CSI is presented in Section
IV. Section V provides the simulation results to verify the
effectiveness of the obtained results in Section III and Section
IV. Finally, conclusion is given in Section VI.

Throughout this paper,  lower-case and upper-case  boldface letters denote vectors and matrices, respectively. ${\mathbb{C}}^{M\times K}$ denotes the $M\times K$ complex space.   ${\bf A}^{\text \dag}$ and ${\bf A}^{-1}$ denote  the Hermitian transpose and the inverse of the matrix ${\bf A}$, respectively. ${\bf I}_{M}$ denotes an $M\times M$ identity matrix. ${\bf 0}_{M\times K}$ denotes an $M\times K$ zero matrix. The $\mathbb{E}_{X}\{\cdot\}$ denotes expectation with respect to the random variable $X$.   A complex Gaussian random vector ${\bf x}$ is denoted as ${\bf x}\sim\mathcal {C}\mathcal {N}(\bar{{\bf x}},{\bf {\Sigma}})$, where the mean vector is $\bar{{\bf x}}$ and the covariance matrix is ${\bf {\Sigma}}$. $\|\cdot\|_{1}$ and $\|\cdot\|_{2}$ denote the 1-norm and 2-norm of a vector, respectively.
${\rm {diag}}\left({\bf a}\right)$ denotes a diagonal matrix where the main diagonal entries are the elements of vector ${\bf a}$. $\Gamma(x)=\int_{0}^{\infty}t^{x-1}e^{-t}\mathrm{d}t$ is the Gamma function. Finally, for given functions $f(L)$ and $g(L)$, notations $\Theta(g(L))=\{f(L):\exists c_1, c_2, L_0>0,\forall L\geq L_0, 0\leq c_1g(L)\leq f(L)\leq c_1g(L)\}$, $O(g(L))=\{f(L):\exists c, L_0>0,\forall L\geq L_0, 0\leq f(L)\leq cg(L)\}$, and $\Omega(g(L))=\{f(L):\exists c,L_0>0,\forall L\geq L_0, 0\leq cg(L)\leq f(L)\}$ mean that $g(L)$ is an asymptotically tight bound, an asymptotic upper bound, and an asymptotic lower bound for $f(L)$, respectively, where the detailed definitions can be found in \cite[Ch. 3.1]{Cormen09}.

\section{System Model}
\begin{figure}[!t]
\centering
\includegraphics[width=2.8in]{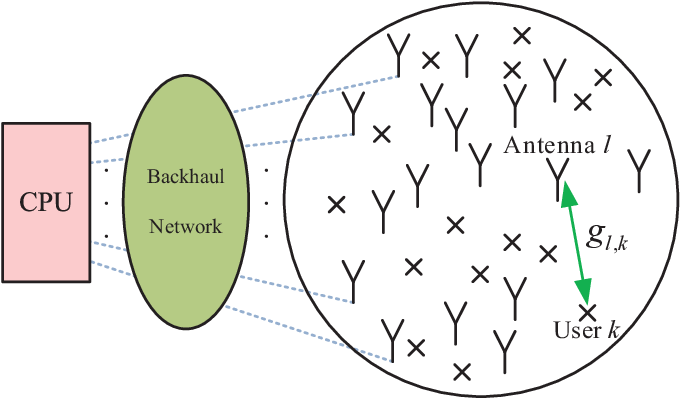}
\caption{A cell-free massive MIMO system with a circular area. X-shaped and Y-shaped icons denote a user and a BS antenna, respectively.}
\label{Fig. 1}
\end{figure}
As shown in Fig. 1, a cell-free massive MIMO system with a circular area has $L$ BS antennas and $K$ single-antenna users ($L\gg K$ and $K$ is finite) within uniform distribution \cite{Ngo17}. All users are served by all BS antennas in the same time-frequency resource simultaneously \cite{Ngo17,Nayebi17,Ngo173}.
All BS antennas are linked to an central processing unit (CPU) based on a backhaul network which is used to perform uploading/downloading  network information, i.e., received signal, CSI, etc. For convenience, we assume that the whole system adopts a time-division duplexing (TDD) protocol, as well as, the BS and all users are perfectly synchronized in each symbol. Without loss of generality, the radius of this circular area is 1.


\subsection{Channel Model}
Let ${\bf g}_{k}$ denote the channel vector  between the user $k$ and all $L$ BS antennas as
\begin{align}
{\bf g}_{k}=\sqrt{{\rm {diag}}\left({\bm \gamma}_{k}\right)}{\bf h}_{k}\in\mathbb{C}^{L\times 1},\tag{1}
\end{align}
where ${\bf h}_{k}=[h_{1,k},\ldots,h_{l,k},\ldots,h_{L,k}]^{\text T}\sim\mathcal {C}\mathcal {N}({\bf 0}_{L\times 1},{\bf I}_{L})$ denotes the $L\times 1$ small-scale fading vector and ${\bm \gamma}_{k}=[\gamma_{1,k},\ldots,\gamma_{l,k},\ldots,\gamma_{L,k}]^{\text T}$ denotes the $L\times 1$ large-scale fading vector. To be more specific, the $l$th entry ${g}_{l,k}$ of the vector ${\bf g}_{k}$ is written as
\begin{align}
{g}_{l,k}={h}_{l,k}\gamma_{l,k}^{\frac{1}{2}}.\tag{2}
\end{align}
For convenience, we  model the large-scale fading coefficient ${\gamma}_{l,k}$ as
\begin{align}
{\gamma}_{l,k}=d_{l,k}^{-{\alpha}},\tag{3}
\end{align}
where $d_{l,k}$ is the distance between the $l$th BS antenna and user $k$, and
$\alpha$ denotes the path-loss factor \cite{Hoydis13,Dai14}. As shown in Fig. 1, ${\bm \gamma}_{k}$ has unequal entries since the user $k$ in the circular area has different access distances to all $L$ BS antennas. Here, a block-fading model is considered as in \cite{Ngo13},  which means that the channels are both constant and frequency flat during a coherence block. Hence, based on slow-varying characteristic of the large-scale fading, and the large-scale fading coefficients stay constant for many coherence intervals, which can be obtained by channel measurement and feedback \cite{Dai14},  without loss of generality, we assume that the large-scale fading information is perfectly available at the BS \cite{Ngo17,Ngo173,Ngo18}.
At the same time, the assumption is taken that each user's channel is  independent from other users' channels.

\subsection{Uplink Data Transmission}
For uplink data transmission, after all $K$ users send uplink data to the BS synchronously, the received signal vector ${\bf y}\in\mathbb{C}^{L \times 1}$ at the BS is modeled as
\begin{align}
{\bf y}&=\sqrt{\rho_u}\sum\limits_{k=1}^{K}{\bf g}_{k}{s}_{k}+{\bf z}\notag\\
&=\sqrt{\rho_u}{\bf G}{\bf s}+{\bf z},\tag{4}
\end{align}
where ${s}_{k}$ is the information-bearing signal transmitted by user $k$, which is the $k$th entry of vector ${\bf s}\in\mathbb{C}^{K\times 1}$ with the conditions $\mathbb{E}_{{\bf s}}\{{\bf s}\}={\bf 0}_{K\times 1}$ and $\mathbb{E}_{{\bf s}}\{{\bf s}{\bf s}^{\dag}\}={\bf I}_{K}$, $\rho_u$ is the transmit power, ${\bf z}\sim\mathcal {C}\mathcal {N}({\bf 0}_{L\times 1},{\bf I}_{L})$ denotes the received noise vector at the BS, and ${\bf G}=[{\bf g}_{1},\ldots,{\bf g}_{k},\ldots,{\bf g}_{K}]\in \mathbb{C}^{L \times K}$ is the channel matrix between all users and all BS antennas.  Moreover, we assume that $\bf z$ is uncorrelated with any user channel and transmitted symbol.


\section{SE Analysis With Perfect CSI}
\subsection{General Achievable Uplink Rate}
We investigate the case that the BS knows the complete channel gain, i.e., $\bf G$.  The received signal $\bf y$ in (4) is separated into $K$ streams by multiplying it with the
linear detector matrix ${\bf A}\in\mathbb{C}^{L\times K}$ which is a function of $\bf G$. The processed signal is given by
\begin{align}
{\bf r}&={\bf A}^{\dag}{\bf y}\in\mathbb{C}^{K\times 1}.\tag{5}
\end{align}
Then, the $k$th entry $r_{k}$ of ${\bf r}$ is modeled as
\begin{align}
{r}_{k}=\sqrt{\rho_u}{\bf a}_{k}^{\dag}{\bf g}_{k}s_{k}+\sqrt{\rho_u}\sum\limits_{n\neq k}^{K}{\bf a}_{k}^{\dag}{\bf g}_{n}s_{n}+{\bf a}_{k}^{\dag}{\bf z},\tag{6}
\end{align}
where ${\bf a}_{k}$ is the $k$th column of ${\bf A}$. Note that the desire signal of user $k$ is  $\sqrt{\rho_u}{\bf a}_{k}^{\dag}{\bf g}_{k}s_{k}$ while the rest of in (6) can be regarded as the interference signal that these two parts are uncorrelated.
By modeling the interference signal as an Gaussian noise, utilizing the worst case technology in \cite[Theorem 1]{Hassibi03}, adopting the standard linear ZF detector ${\bf A}={\bf G}_{}\left({\bf G}_{}^{\dag}{\bf G}_{}\right)^{-1}$, and  normalizing the total system bandwidth into unity,
the achievable uplink rate (i.e., SE) of user $k$, in units of bit/s/Hz, is given by \cite[Eq. (18)]{Ngo13}
\begin{align}
R_{k}=\mathbb{E}_{\bf H}\left\{\log_2\left(1+\frac{\rho_u}{\|{\bf a}_{k}\|_2^{2}}\right)\right\},\tag{7}
\end{align}
where
\begin{align}
{\bf H}\triangleq[{\bf h}_{1},\ldots,{\bf h}_{k},\ldots,{\bf h}_{K}]\in \mathbb{C}^{L \times K}.\tag{8}
\end{align}
It is important to note that $R_k$ is determined by the users' large-scale fading coefficients. Hence, to analyze the average SE performance and study the asymptotic performance  of (7), it is necessary to evaluate the achievable uplink rate by averaging out the large-scale fading. We define the following rate metric as \cite[Eq. (12)]{Wang15} and \cite[Eq. (4)]{Matthaiou13}
\begin{align}
\bar{R}_{k}=\mathbb{E}_{\bm \Upsilon}
\left\{R_{k}\right\},\tag{9}
\end{align}
where
\begin{align}
{\bm \Upsilon}\triangleq\left[{\bm \gamma}_{1},\ldots,{\bm \gamma}_{k},\ldots,{\bm \gamma}_{K}\right]\in\mathbb{C}^{L\times K}.\tag{10}
\end{align}

\subsection{Upper Bound}

{\emph {Theorem 1:}} An upper bound $R_{k}^{\text {ub}}$ of $R_{k}$
has an approximate closed-form expression when $L$ is large, which is given by
\begin{align}
R_{k}^{\text {ub}}\approx\widetilde{{R}_{k}^{\text {ub}}}\triangleq\log_2\left(1+\rho_{u}\sum\limits_{\tilde{l}\in{\mathcal {L}}_{k}}\gamma_{\tilde{l},k}\right),\tag{11}
\end{align}
where ${\mathcal {L}}_{k}={{\mathcal {L}}/{\mathcal {A}}_{k}}$, ${\mathcal {L}}=\{l| \forall l=1,\ldots, L\}$, and ${\mathcal {A}}_{k}$ is defined as
\begin{align}
{\mathcal {A}}_{k}\triangleq{\text {Unique}}\left(\left\{l_{n}^{\star}=\arg\max\limits_{l}\gamma_{l,n}\bigg| \forall n\neq k\right\}\right),\tag{12}
\end{align}
and ${\text {Unique}}\left(\mathcal {T}\right)$ returns the same values as in set $\mathcal {T}$ but with no repetitions.

{\emph {Proof:}} See Appendix B. \hfill\rule{3mm}{3mm}

It is important to note that the result in {\emph {Theorem 1}} does not depend on the small-scale fading and is constituted by a series of large-scale fading coefficients, which change less often than the practical channel. Hence, the formula in {\emph {Theorem 1}} only needs to be calculated when the large-scale fading coefficients change. Also, the obtained close-form expression enables  efficient evaluation of the achievable uplink rate and providing key insights about how the achievable uplink rate is affected by the system parameters. Specifically, based on (11) and the special structure of (12), it is found that, for the approximation of $R_{k}$, at most $K-1\geq|{\mathcal {A}}_{k}|$ large-scale coefficients become useless after the BS processes the received signal vector $\bf y$ in (4) by the ZF detector.
Intuitively, the essence of the user $k$'s detector ${\bf a}_{k}$ is to project the signal vector ${\bf y}$ onto the null space of the interference space spanned by $\{{\bf g}_{n}|\forall n\neq k\}$. Also, since the BS antenna $l_{n}^{\star}$ is the closest one to the user $n$, the corresponding large-scale fading coefficient is much larger than $\gamma_{l,n} (l\neq l_{n}^{\star})$, hence $\gamma_{l_{n}^{\star},n}$ becomes significant and $g_{l_{n}^{\star},n}$ is the principal element of ${\bf g}_{n}$, therefore, the $l_{n}^{\star}$th component $g_{l_{n}^{\star},k}$ of ${\bf g}_{k}$ is approximately eliminated. In summary, from both theorem analysis and physics intuition perspectives, the obtained approximation of $R_{k}$ in {\emph {Theorem 1}} not only has its rationality, but also this closed-form expression  allows us to pursuit the benefit of the important system design. In spite of the large $L$ assumption in {\emph {Theorem 1}}, it is worth mentioning that this result is still valid with a not-so-large number of antennas and is larger than the simulated rate in section V.
Moreover, motivated by the previous work \cite[Remark 5]{Ngo17}, it is found that the obtained approximation can reduce to the exact upper bound obtained in conventional massive MIMO systems based on properly utilizing several conditions. The detailed explanation is as follows. First, when the number of BS antennas $L$ becomes infinite, based on the special structure of ${\mathcal {A}}_{k}$ and ${\mathcal {L}}_{k}$, the cardinal of the set ${\mathcal {A}}_{k}$ asymptotically approaches to $K-1$ and hence $|{\mathcal {L}}_{k}|\rightarrow L-K+1$. Then, by making the antennas belong to the set ${\mathcal {L}}_{k}$ at the same location, i.e., $\forall \tilde{l}\in{\mathcal {L}}_{k}, \gamma_{\tilde{l},k}=\gamma_{k}$, $\widetilde{{R}_{k}^{\text {ub}}}$ finally becomes
\begin{align}
\widetilde{{R}_{k}^{\text {ub}}}\rightarrow
\log_2\left(1+\rho_u\left(L-K+1\right)\gamma_{k}\right).\tag{13}
\end{align}
Note that (13) is the exact upper bound of the achievable uplink rate  with ZF detector and perfect CSI
in a conventional massive MIMO system given by \cite[Theorem 3]{QiZhang15} if the inter-cell interference is ignored. In summary, {\emph {Theorem 1}} provides a very generic expression (11) to approximate the limit system performance.

It is important to note that different large-scale fading generates different (11). However, the exact expression of  averaging out the large-scale fading in $\widetilde{{R}_{k}^{\text {ub}}}$ is hard to be obtained since the distribution of the access distance between the user and BS antenna is complicated as shown in {\emph {Lemma 1}}  of the Appendix A. The following corollary will be used for investigating the asymptotic performance of the approximation of $R_{k}$ by averaging out the large-scale fading.

{\emph {Corollary 1:}} An upper bound $\bar{R}_{k}^{\text {ub}}$ of $\bar{R}_{k}$ has the following approximation
\begin{align}
\bar{R}_{k}^{\text {ub}}\triangleq\mathbb{E}_{\bm \Upsilon}
\left\{{{R}_{k}^{\text {ub}}}\right\}\approx\widetilde{\bar{R}_{k}^{\text {ub}}}\triangleq\mathbb{E}_{\bm \Upsilon}
\left\{\widetilde{{R}_{k}^{\text {ub}}}\right\}.\tag{14}
\end{align}
When $L\rightarrow\infty$, $\widetilde{\bar{R}_{k}^{\text {ub}}}=\Omega(\frac{\alpha}{2}\log_2L)$.


{\emph {Proof:}}  See Appendix C. \hfill\rule{3mm}{3mm}

It is interesting to note from {\emph {Corollary 1}} that the growth rate of the approximation of achievable uplink rate $\bar{R}_{k}$, by averaging out the large-scale fading, is not only the logarithmic function of the number of BS antennas $L$ but also the linear function of the path-loss factor $\alpha$.
The underlying reason behind this phenomenon is that the different access distances from the user to the massive distributed BS antennas offer extra {\emph {distance diversity}} compared with the case when all massive BS antennas are placed together.


\subsection{Lower Bound}

{\emph {Theorem 2:}} A lower bound $R_{k}^{\text {lb}}$ of $R_{k}$
has an approximate closed-form expression when $L$ is large, which is given by
\begin{align}
R_{k}^{\text {lb}}\approx\widetilde{{R}_{k}^{\text {lb}}}
\triangleq\log_2\left(1+\rho_{u}\Phi_{k}\left(\Psi_{k}-1\right)\right),\tag{15}
\end{align}
where $\Phi_{k}$ and $\Psi_{k}$ are defined as
\begin{align}
&\Phi_{k}\triangleq
\frac{\sum\limits_{\tilde{l}\in{\mathcal {L}}_{k}}\gamma_{\tilde{l},k}^2}
{\sum\limits_{\tilde{l}\in{\mathcal {L}}_{k}}\gamma_{\tilde{l},k}},\tag{16}\\
&\Psi_{k}\triangleq
\frac{\left(\sum\limits_{\tilde{l}\in{\mathcal {L}}_{k}}\gamma_{\tilde{l},k}\right)^2}
{\sum\limits_{\tilde{l}\in{\mathcal {L}}_{k}}\gamma_{\tilde{l},k}^2},\tag{17}
\end{align}
respectively.

{\emph {Proof:}}  See Appendix D. \hfill\rule{3mm}{3mm}

It is important to note that the result in ${\emph {Theorem 2}}$ presents the simple analytical investigation of the approximation of the achievable uplink rate $R_k$ for the case of each BS antennas is distributed. It is worth noting that,  as $L\rightarrow\infty$, by adopting the same assumption for obtaining (13), $\widetilde{\hat{R}_{k}^{\text {lb}}}$ becomes
\begin{align}
\widetilde{{R}_{k}^{\text {lb}}}\rightarrow
\log_2\left(1+\rho_u\left(L-K\right)\gamma_{k}\right).\tag{18}
\end{align}
It is well known that (18) is the exact lower bound of the achievable uplink rate with ZF detector and perfect CSI in a conventional massive MIMO system given by \cite[Proposition 3]{Ngo13}. Based on this fact, {\emph {Theorem 2}} provides an useful formula to assess the achievable uplink rate performance for the general communication system. Moreover, it is obvious that $\widetilde{{R}_{k}^{\text {lb}}}$ is always smaller than $\widetilde{{R}_{k}^{\text {ub}}}$, which implies the validity of {\emph {Theorem 2}}. Next, it is interesting to consider the scaling behavior of the obtained approximation by averaging out the large-scale fading.

{\emph {Corollary 2:}} An lower bound $\bar{R}_{k}^{\text {lb}}$ of $\bar{R}_{k}$ has the following approximation
\begin{align}
\bar{R}_{k}^{\text {lb}}\triangleq\mathbb{E}_{\bm \Upsilon}
\left\{{{R}_{k}^{\text {lb}}}\right\}\approx\widetilde{\bar{R}_{k}^{\text {lb}}}\triangleq\mathbb{E}_{\bm \Upsilon}
\left\{\widetilde{{R}_{k}^{\text {lb}}}\right\}.\tag{19}
\end{align}
When $L\rightarrow\infty$, $\widetilde{\bar{R}_{k}^{\text {lb}}}=\Omega(\frac{\alpha}{2}\log_2L)$.

{\emph {Proof:}}  See Appendix E. \hfill\rule{3mm}{3mm}

Note that an analogous result has  been derived in {\emph {Corollary 1}}. To gain more insights, the conclusions in  {\emph {Corollary 1}} and  {\emph {Corollary 2}} imply that the obtained two approximations, where line at either end of the achievable uplink rate with perfect CSI, have the same asymptotic lower bound $\frac{\alpha}{2}\log_2L$ and the path-loss factor $\alpha$ is always larger than 2 unless in the case of free space propagation. Interestingly, by placing all antennas at the area centre and averaging out the large-scale fading, based on (13) and (18), it is found that the achievable uplink rate of ZF detector with perfect CSI in the conventional  massive MIMO system is  $\Theta(\log_2L)$ which means the growth rate is no further than  $\log_2L$.  In brief, cell-free massive MIMO systems promise better achievable uplink rate performance than the conventional massive MIMO systems based on the conditions of the perfect CSI assumption and ZF detector. In the following section, the in-depth studies have been made on the achievable uplink rate analysis with imperfect CSI case.

\section{SE Analysis With Imperfect CSI}

\subsection{General Achievable Uplink Rate}
In real communication system, a key component is to obtain the CSI between the users and the BS antennas. In TDD model, all users firstly send uplink pilot sequences to the BS which knows the exact pilot sequence information. Then, the BS utilizes these pilot sequences and the statistical information to estimate the channels of all users. For the sake of analyzing the SE performance in a simplified manner, the typical  minimum mean squared error  estimation method \cite{Kay93} and the orthogonal pilot sequences \cite{Ngo13} are adopted. The total channel $\bf G$ is estimated as
\begin{align}
\hat{\bf G}\triangleq\left[\hat{\bf g}_{1},\ldots,\hat{\bf g}_{k},\ldots,\hat{\bf g}_{K}\right]\in\mathbb{C}^{L\times K},\tag{20}
\end{align}
where $\hat{\bf g}_{k}=[\hat{g}_{1,k},\ldots,\hat{g}_{l,k},\ldots,\hat{g}_{L,k}]^{\text T}\in\mathbb{C}^{L\times 1}$ is the estimated channel vector of ${\bf g}_{k}$. Besides, $\forall k$, $\hat{\bf g}_{k}$ and the estimate error vector $\tilde{\bf g}_{k}\triangleq{\bf g}_{k}-\hat{\bf g}_{k}$ follow
\begin{align}
\hat{\bf g}_{k}&\sim\mathcal {C}\mathcal {N}({\bf 0}_{L\times 1}, {\rm {diag}}(\hat{\gamma}_{1,k},\ldots,\hat{\gamma}_{l,k},\ldots,\hat{\gamma}_{L,k})),\tag{21}\\
\tilde{\bf g}_{k}&\sim\mathcal {C}\mathcal {N}({\bf 0}_{L\times 1}, {\rm {diag}}(\tilde{\gamma}_{1,k},\ldots,\tilde{\gamma}_{l,k},\ldots,\tilde{\gamma}_{L,k})),\tag{22}
\end{align}
and satisfy $\mathbb{E}_{{\bf h}_{k}}\{\hat{\bf g}_{k}^{\dag}\tilde{\bf g}_{k}\}=0$, where
\begin{align}
\hat{\gamma}_{l,k}&\triangleq\frac{\rho_p{\gamma}_{l,k}}{\rho_p{\gamma}_{l,k}+1}{\gamma}_{l,k},\tag{23}\\
\tilde{\gamma}_{l,k}&\triangleq\frac{1}{\rho_p{\gamma}_{l,k}+1}{\gamma}_{l,k},\tag{24}
\end{align}
and $\rho_p$ denotes the pilot sequence power. Similar as the perfect CSI case, we multiply (4) with the linear detector $\hat{\bf A}$ constituted by $\hat{\bf G}$  as follow
\begin{align}
\hat{\bf r}_{}=\hat{\bf A}^{\dag}{\bf y}_{}.\tag{25}
\end{align}
Then, the $k$th entry $\hat{r}_{k}$ of $\hat{\bf r}_{}$ is given by
\begin{align}
\hat{r}_{k}=\sqrt{\rho_u}\hat{\bf a}_{k}^{\dag}{\bf g}_{k}s_{k}+\sqrt{\rho_u}\sum\limits_{n\neq k}^{K}\hat{\bf a}_{k}^{\dag}{\bf g}_{n}s_{n}+\hat{\bf a}_{k}^{\dag}{\bf z},\tag{26}
\end{align}
where $\hat{\bf a}_{k}$ is the $k$th column of $\hat{\bf A}_{}$.
Since the  BS estimates $\bf G$ as $\hat{\bf G}$, $\hat{r}_{k}$ is rewritten as
\begin{align}
\hat{r}_{k}&\!=\!\sqrt{\rho_u}\hat{\bf a}_{k}^{\dag}\hat{\bf g}_{k}s_{k}
\!+\!\sqrt{\rho_u}\!\sum\limits_{n\neq k}^{K}\hat{\bf a}_{k}^{\dag}\hat{\bf g}_{n}s_{n}
\!+\!\sqrt{\rho_u}\!\sum\limits_{n=1}^{K}\hat{\bf a}_{k}^{\dag}\tilde{\bf g}_{n}s_{n}
\!+\!\hat{\bf a}_{k}^{\dag}{\bf z}.\tag{27}
\end{align}
Then, by  utilizing the worst-case technique of Gaussian noise from \cite[Theorem 1]{Hassibi03},
, adopting the standard ZF detector $\hat{\bf A}=\hat{\bf G}_{}\left(\hat{\bf G}_{}^{\dag}\hat{\bf G}_{}\right)^{-1}$, and normalizing
the total system bandwidth into unity, the achievable uplink rate of the $k$th user by averaging out the small-scale fading $\bf H$ is given as
\begin{align}
\hat{R}_{{k}}=\mathbb{E}_{\bf H}\left\{\log_2\left(1+\frac{\rho_u}{\rho_u\sum\limits_{l=1}^{L}|\hat{ a}_{l,k}|^2\sum\limits_{n=1}^{K}\tilde{\gamma}_{l,n}+\|\hat{\bf a}_{k}\|_2^{2}}\right)\right\},\tag{28}
\end{align}
where  $\hat{ a}_{l,k}$ is the $l$th entry of $\hat{\bf a}_{k}$. Note that with high $\rho_p$, hence $\tilde{\gamma}_{l,n}\rightarrow0 (\forall l,n)$, $\hat{\bf a}_{k}\rightarrow{\bf a}_{k}$, and $\hat{R}_{{k}}$ becomes the perfect CSI case as in (7).\footnote{For the sake of characterizing the variation of the achievable uplink rate from the imperfect CSI case to the perfect CSI case, the channel estimation overhead in the coherence time interval does not take into account as in \cite{Ngo13}.} Moreover, by following the methodology in the perfect CSI case, the achievable uplink rate of the $k$th user by averaging out both small-scale fading and large-scale fading is given by
\begin{align}
\bar{{\hat R}}_{k}=\mathbb{E}_{\bm \Upsilon}
\left\{{\hat R}_{k}\right\}.\tag{29}
\end{align}

\subsection{Upper Bound}

{\emph {Theorem 3:}}  An upper bound $\hat{R}_{k}^{\text {ub}}$ of $\hat{R}_{k}$
has an approximate closed-form expression when $L$ is large, which is given by
\vspace{-2mm}
\begin{align}
\hat{R}_{k}^{\text {ub}}\approx\widetilde{\hat{R}_{k}^{\text {ub}}}\triangleq\log_2\left(1
+\frac{\rho_{u}}{\rho_u\tilde{\gamma}_{\min}+1}\sum\limits_{\tilde{l}\in{{\mathcal {L}}}_{k}}\hat{\gamma}_{\tilde{l},k}\right),\tag{30}
\end{align}
where  $\tilde{\gamma}_{\min}$ is defined as
\vspace{-3mm}
\begin{align}
\tilde{\gamma}_{\min}\triangleq\min_{l}
\sum\limits_{n=1}^{K}\tilde{\gamma}_{l,n}.\tag{31}
\end{align}

{\emph {Proof:}} Based on the Jensen's inequality and the definition of (31), (28) is upper bounded by
\vspace{-2mm}
\begin{align}
\hat{R}_{k}&\leq
\log_2\left(1+\frac{\rho_{u}}{\rho_u\tilde{\gamma}_{\min}+1}
\mathbb{E}_{\bf H}\left\{\frac{1}{\|\hat{\bf a}_{k}\|_2^2}\right\}
\right)\triangleq \hat{R}_{k}^{\text {ub}},\tag{32}
\end{align}
where the ``$=$" in ``$\leq$" is reached when both $\frac{1}{\|\hat{\bf a}_{k}\|_2^2}$ is a constant and $\sum\limits_{n=1}^{K}\tilde{\gamma}_{l_1,n}\!\!=\!\!\sum\limits_{n=1}^{K}\tilde{\gamma}_{l_2,n},\forall l_1\neq l_2$. The following key issue is to obtain the distribution of $\frac{1}{\|\hat{\bf a}_{k}\|_2^2}$. By applying the methodology of the proof of {\emph {Theorem 1}}, we can reach the final result after some basic algebraic manipulations.

\hfill\rule{3mm}{3mm}


Note that the approximate expression in  {\emph {Theorem 3}} can be easily evaluated since it primarily involves simple large-scale fading coefficients, uplink data power, and a minimum value which depends on the variance of the channel estimation error, as well as, the standard logarithmic function base 2. Moreover,  it is obvious that $\widetilde{\hat{R}_{k}^{\text {ub}}}$ is always smaller than the perfect CSI case $\widetilde{{R}_{k}^{\text {ub}}}$ in (11) and satisfies
\begin{align}
\lim_{\rho_p\rightarrow\infty}\widetilde{\hat{R}_{k}^{\text {ub}}}
=\widetilde{{R}_{k}^{\text {ub}}}.\tag{33}
\end{align}
We note that the conclusion of {\emph {Theorem 3}} gives another universal formula for the approximation of achievable uplink rate. Moreover, when $L\rightarrow\infty$, by utilizing the methodology of obtaining (13), $\widetilde{\hat{R}_{k}^{\text {ub}}}$ becomes
\begin{align}
\widetilde{\hat{R}_{k}^{\text {ub}}}\rightarrow
\log_2\left(1+
\frac{\rho_u\left(L-K+1\right)}
{\rho_u\sum\limits_{n=1}^{K}{\frac{{\gamma}_{n}}{\rho_p{\gamma}_{n}+1}}+1}\!\times\!
\frac{\rho_p{\gamma}_{k}}{\rho_p{\gamma}_{k}+1}{\gamma}_{k}
\right).\tag{34}
\end{align}
An interesting phenomenon is found that (34) is the exact upper bound of the achievable uplink rate with ZF detector and imperfect CSI in a  conventional massive MIMO system given by the $k$th component in \cite[Theorem 1]{Wang172} when the out-of-cell interference is ignored.  Hence, (34) is a special case of (30) if  all the BS antennas are located at the same position.

Similar as the perfect CSI case, the following corollary investigates the  asymptotic performance of ${\hat{R}_{k}^{\text {ub}}}$ by averaging out the large-scale fading.

{\emph {Corollary 3:}} An upper bound $\bar{\hat R}_{k}^{\text {ub}}$ of $\bar{\hat R}_{k}$ has the following approximation
\begin{align}
\bar{\hat R}_{k}^{\text {ub}}\triangleq\mathbb{E}_{\bm \Upsilon}
\left\{{\hat{R}_{k}^{\text {ub}}}\right\}\approx\widetilde{\bar{\hat R}_{k}^{\text {ub}}}\triangleq\mathbb{E}_{\bm \Upsilon}
\left\{\widetilde{\hat{R}_{k}^{\text {ub}}}\right\}.\tag{35}
\end{align}
When $L\rightarrow\infty$, $\widetilde{\bar{\hat R}_{k}^{\text {ub}}}=\Omega(\frac{\alpha}{2}\log_2L)$.

{\emph {Proof:}} The proof is similar as in {\emph {Corollary 1}} and thus omitted.  \hfill\rule{3mm}{3mm}

Together with {\emph {Corollary 1}}, {\emph {Corollary 3}} indicates that if the uplink data power $\rho_u$, pilot sequence power $\rho_p$, the number of users $K$, and the path-loss factor $\alpha$ are kept fixed, as well as, the number of BS antennas $L$ is increased, then, the result of scaling behavior is the same  as the perfect CSI case.  In spite of this, it is worth noting that the gap between the two approximations by averaging out the large-scale fading for the perfect CSI case and imperfect CSI is enlarged when the pilot sequence power $\rho_p$ is cut down.

\vspace{-2mm}
\subsection{Lower Bound}

{\emph {Theorem 4:}} A lower bound ${\hat R}_{k}^{\text {lb}}$ of ${\hat R}_{k}$
has an approximate closed-form expression when $L$ is large, which is given by
\begin{align}
{\hat R}_{k}^{\text {lb}}\approx\widetilde{{\hat R}_{k}^{\text {lb}}}
\triangleq\log_2\left(1+
\frac{\rho_{u}}{\rho_u\tilde{\gamma}_{\max}+1}
{\hat \Phi}_{k}\left({\hat \Psi}_{k}-1\right)\right),\tag{36}
\end{align}
where $\tilde{\gamma}_{\max}$, ${\hat \Phi}_{k}$, and ${\hat \Psi}_{k}$ are defined as
\begin{align}
\tilde{\gamma}_{\max}&\triangleq\max_{l}\sum\limits_{n=1}^{K}\tilde{\gamma}_{l,n},\tag{37}\\
{\hat \Phi}_{k}&\triangleq
\frac{\sum\limits_{\tilde{l}\in{\mathcal {L}}_{k}}{\hat \gamma}_{\tilde{l},k}^2}
{\sum\limits_{\tilde{l}\in{\mathcal {L}}_{k}}{\hat \gamma}_{\tilde{l},k}},\tag{38}\\
{\hat \Psi}_{k}&\triangleq
\frac{\left(\sum\limits_{\tilde{l}\in{\mathcal {L}}_{k}}{\hat \gamma}_{\tilde{l},k}\right)^2}
{\sum\limits_{\tilde{l}\in{\mathcal {L}}_{k}}{\hat \gamma}_{\tilde{l},k}^2},\tag{39}
\end{align}
respectively.

{\emph {Proof:}}
Based on the Jensen's inequality and the definition of (37), (28) is lower bounded by
\begin{align}
\hat{R}_{k}&\geq
\log_2\left(1+\frac{\rho_{u}}{\left(\rho_u\tilde{\gamma}_{\max}+1\right)\mathbb{E}_{\bf H}\left\{{\|\hat{\bf a}_{k}\|_2^2}\right\}}
\right)\triangleq \hat{R}_{k}^{\text {lb}},\tag{40}
\end{align}
where the ``$=$" in ``$\geq$" is reached when both ${\|\hat{\bf a}_{k}\|_2^2}$ is a constant and $\sum\limits_{n=1}^{K}\tilde{\gamma}_{l_1,n}\!\!=\!\!\sum\limits_{n=1}^{K}\tilde{\gamma}_{l_2,n},\forall l_1\neq l_2$.
The following key issue is to obtain the distribution of ${\|\hat{\bf a}_{k}\|_2^2}$.
By applying the methodology of the proof of {\emph {Theorem 2}}, we can reach the final result after some basic algebraic manipulations.

\hfill\rule{3mm}{3mm}


Note that the expression in {\emph {Theorem 4}} can be easily evaluated since it involves the pilot sequence power, uplink data power, and large-scale fading coefficients, as well as, the antenna index set ${\mathcal {L}}_{k}$.  Interestingly, from (36), it is shown that the approximation of the achievable uplink rate with imperfect CSI is similar as the perfect CSI in {\emph {Theorem 2}}, while the channel estimate error is involved that makes a little difference. Moreover, when the pilot sequence power $\rho_p$ becomes infinite, based on (23) and (24), by taking the similar methodology of obtaining (33),
 $\tilde{\gamma}_{\max}\rightarrow0$,  ${\hat \Phi}_{k}\rightarrow{\Phi}_{k}$, ${\hat \Psi}_{k}\rightarrow{\Psi}_{k}$, and hence
\begin{align}
\lim_{\rho_p\rightarrow\infty}\widetilde{\hat{R}_{k}^{\text {lb}}}=\widetilde{{R}_{k}^{\text {lb}}}.\tag{41}
\end{align}
Moreover, with finite $\rho_p$, based on the monotonic property from the proof of {\emph {Corollary 2}} in Appendix E, after some algebraic manipulation, it can be shown that $\widetilde{\hat{R}_{k}^{\text {lb}}}<\widetilde{{R}_{k}^{\text {lb}}}$. In other words, $\widetilde{\hat{R}_{k}^{\text {lb}}}$ is very generic and can be regarded as a considered model to investigate the performance of achievable uplink rate for the imperfect CSI case and its special case perfect CSI. Moreover, it is easy to show that $\widetilde{\hat{R}_{k}^{\text {lb}}}<\widetilde{\hat{R}_{k}^{\text {ub}}}$, which implies the validity of {\emph {Theorem 4}}.
Particularly, when $L\rightarrow\infty$, by taking the same assumption for obtaining (13), $\widetilde{\hat{R}_{k}^{\text {lb}}}$ becomes
\begin{align}
\widetilde{\hat{R}_{k}^{\text {lb}}}\rightarrow
\log_2\left(1+
\frac{\rho_u\left(L-K\right)}
{\rho_u\sum\limits_{n=1}^{K}{\frac{{\gamma}_{n}}{\rho_p{\gamma}_{n}+1}}+1}\!\times\!
\frac{\rho_p{\gamma}_{k}}{\rho_p{\gamma}_{k}+1}{\gamma}_{k}
\right).\tag{42}
\end{align}
It is important to note that (42) is the exact lower bound of the
achievable uplink rate with ZF detector and imperfect CSI
in the conventional massive MIMO system given by \cite[Proposition 7]{Ngo13}. Hence, $\widetilde{\hat{R}_{k}^{\text {lb}}}$ can be emerged as a promising, tractable, and effective expression to reap both benefits of the rate analysis of cell-free massive MIMO systems and conventional massive MIMO systems.

The following corollary presents an analysis for $\widetilde{\hat{R}_{k}^{\text {lb}}}$ by averaging out the large-scale fading coefficients.

{\emph {Corollary 4:}} An lower bound $\bar{\hat R}_{k}^{\text {lb}}$ of $\bar{\hat R}_{k}$ has the following approximation
\begin{align}
\bar{\hat R}_{k}^{\text {lb}}\triangleq\mathbb{E}_{\bm \Upsilon}
\left\{{\hat{R}_{k}^{\text {lb}}}\right\}\approx\widetilde{\bar{\hat R}_{k}^{\text {lb}}}\triangleq\mathbb{E}_{\bm \Upsilon}
\left\{\widetilde{\hat{R}_{k}^{\text {lb}}}\right\}.\tag{43}
\end{align}
When $L\rightarrow\infty$, $\widetilde{\bar{\hat R}_{k}^{\text {lb}}}=\Omega(\frac{\alpha}{2}\log_2L)$.

{\emph {Proof:}} The proof is similar as in {\emph {Corollary 2}} and thus omitted. \hfill\rule{3mm}{3mm}

%
%
%

Interestingly, the comparison of {\emph {Corollary 3}} and {\emph {Corollary 4}} reveals that the obtained two approximations of the achievable uplink rate with ZF detector and imperfect CSI, have the same asymptotic lower bound $\frac{\alpha}{2}\log_2 L$, where the similar results can be found in {\emph {Corollary 1}} and {\emph {Corollary 2}} for the perfect CSI case. We note that the asymptotic results in {\emph {Corollary 3}} and {\emph {Corollary 4}} are not affected by the pilot sequence power $\rho_p$. It means the degree of the channel estimation error introduced by the phase of channel estimation does merely change the deviation between the imperfect CSI case and perfect CSI case, since the channel estimation does not affect the asymptotic result and only has an adverse impact on the achievable uplink rate. Apart from that, in conventional massive MIMO systems with imperfect CSI,  (34) and  (42) show that the growth rate of achievable uplink rate is just not more than $\log_2 L$ which is same as the perfect CSI case. Hence, cell-free massive MIMO systems have the potential of high growth rate compared with the conventional massive MIMO systems in imperfect CSI.


\section{Numerical Results}
In this section, the validation of the theoretical analysis and asymptotic conclusion of the achievable uplink rate in Section III and Section IV is conducted via numerical simulation. The $L$ BS antennas and $K$ users are uniformly distributed at a circle area with radius one. Moreover, the impact of  the pilot sequence power $\rho_p$ is considered to investigate the achievable uplink rate performance. Specifically, by employing the methodology of \cite{Dai14,Ngo13}, we choose the path-loss factor $\alpha=4$ and the number of users $K=10$. Note that since  we assume the noise variance is 1, $\rho_u$ ($\rho_p$) can be interpreted as the transmit uplink data signal (pilot sequence signal) to noise ratio and, thus, can be expressed in dB.
The simulation results are obtained by randomly producing 30 realizations of all users' positions and 30 realizations of antennas positions in this circular area, respectively. In other words, $30\times30=900$ realizations of the large-scale fading are obtained for numerical simulation. Also, 200 realizations of the small-scale fading channels are randomly produced for each large-scale fading realization. For convenience, in this cell-free massive MIMO system, we define the metrics called ``Average achievable uplink rate"  for the perfect CSI case and imperfect CSI case, which are given by
\begin{align}
\bar{R}&\triangleq\frac{1}{K}\sum\limits_{k=1}^{K}\bar{R}_{k},\tag{44}\\
\bar{\hat R}&\triangleq\frac{1}{K}\sum\limits_{k=1}^{K}\bar{\hat R}_{k},\tag{45}
\end{align}
respectively. Besides, $\bar{R}_{k}$ and $\bar{\hat R}_{k}$ have been defined as in (9) and (29), respectively. Moreover, for further comparison, the conventional massive MIMO system is considered in this section, where
the parameter setting and metric are the same as in the cell-free massive MIMO system, except for all the BS antennas are placed and fixed at this circular area centre.

\begin{figure*}[!t]
\centering
\subfigure[Perfect CSI case] { \label{fig:(a)}
\includegraphics[width=2.8in]{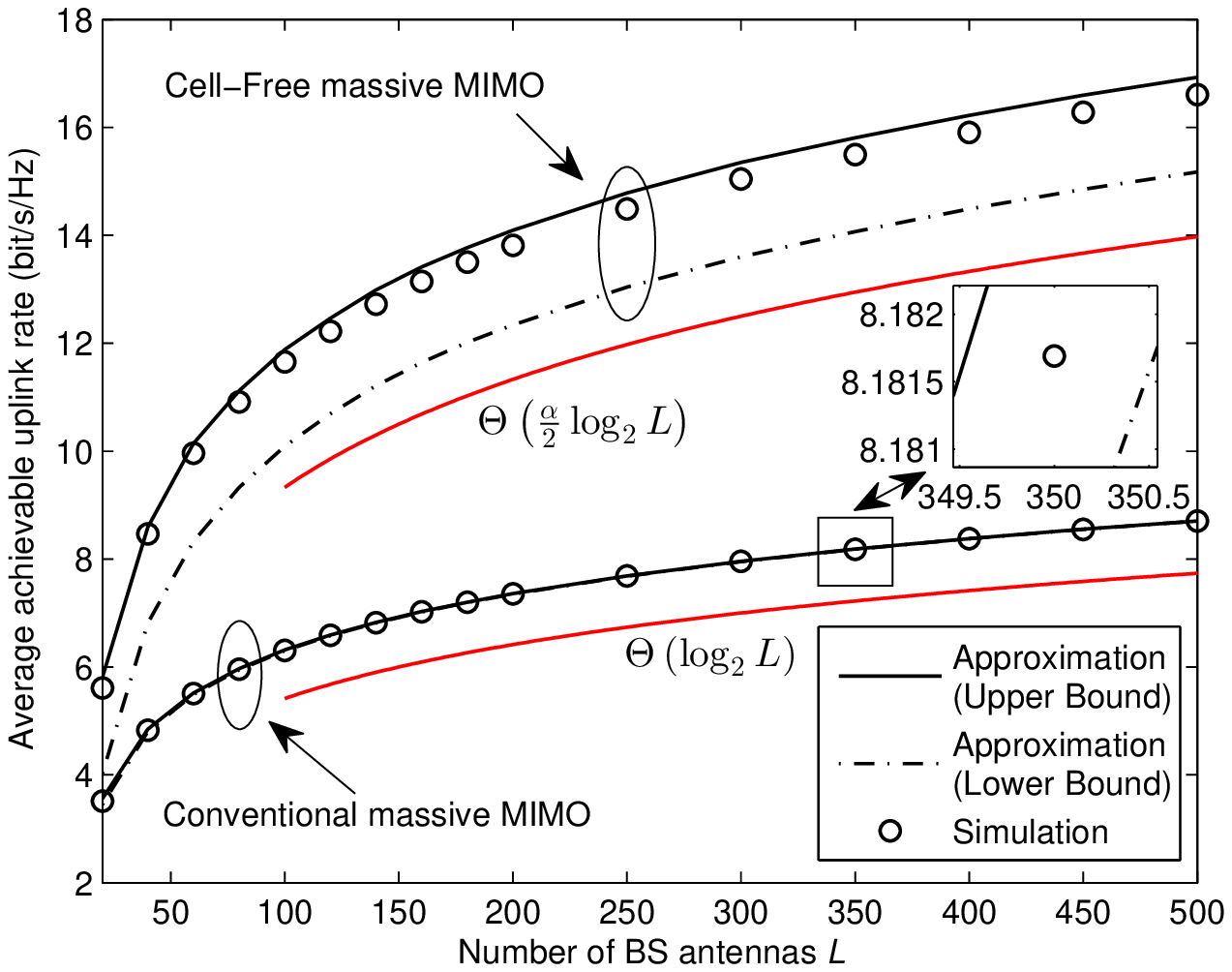}
}
\subfigure[Imperfect CSI case] {\label{fig:(b)}
\includegraphics[width=2.8in]{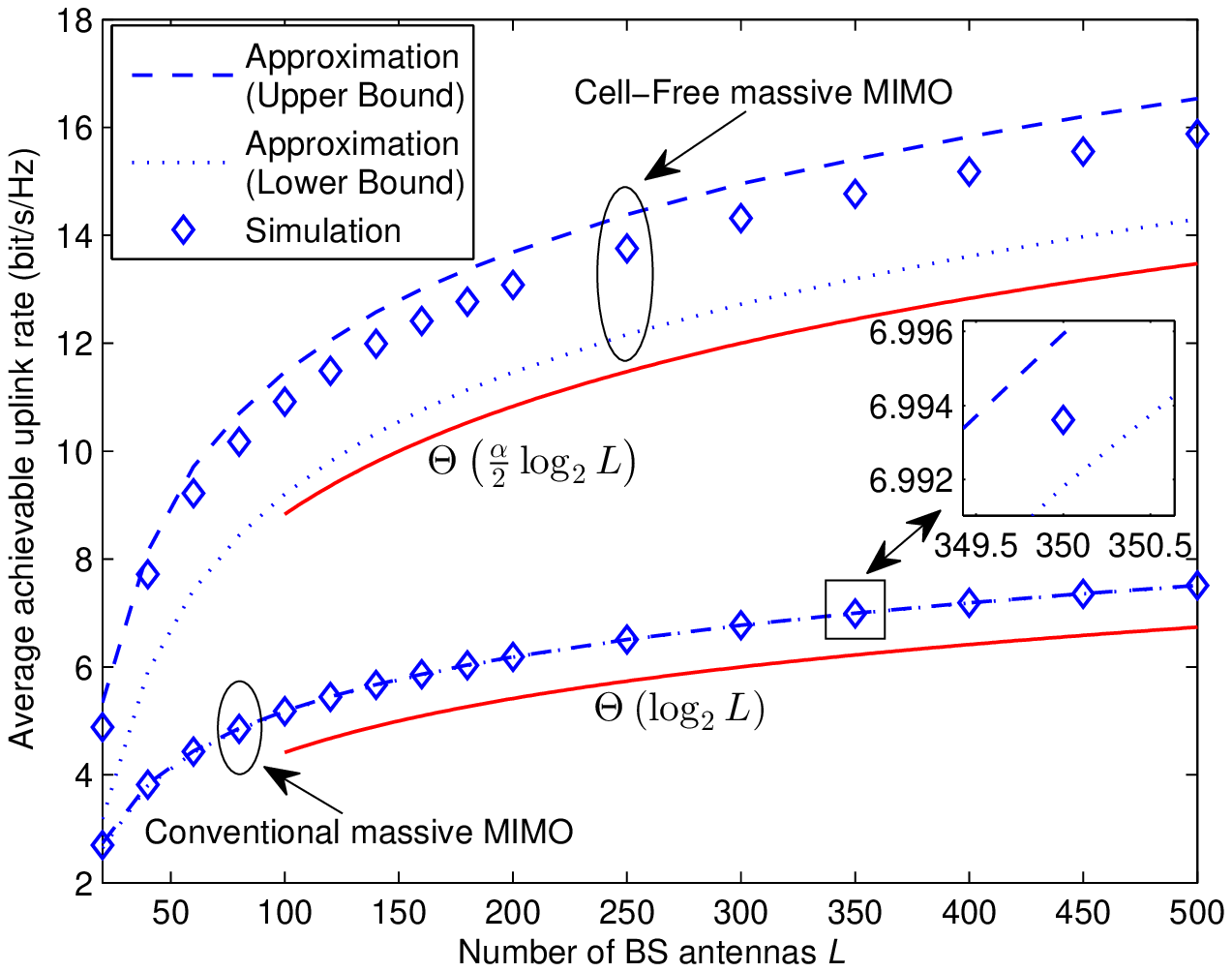}
}
\caption{The average achievable uplink rate when $\rho_u=-10$dB  for cell-free and conventional massive MIMO systems with the perfect CSI and imperfect CSI ($\rho_p=0$dB).}
\label{fig2}
\end{figure*}

Fig. 2  gives the approximations (or bounds)  and  Monte-Carlo simulated average achievable uplink rate for cell-free massive MIMO system and conventional massive MIMO system.\footnote{Note that the obtained approximations in cell-free massive MIMO system and the upper and lower bounds in conventional massive MIMO system are showed in Fig. 2-4. For convenience, we use the formats ``Approximation (Upper Bound)" and ``Approximation (Lower Bound)" in Fig. 2-4.}
Here, due to the orthogonal pilot sequences are adopted in Section IV.A that the pilot sequence length should be larger than or equal to the number of users, for convenience, the uplink data power $\rho_u$ is set to $-10$dB and the pilot  sequence power is set to $\rho_p=K\rho_u=0$dB.


For the perfect CSI case of Fig. 2 (a), with the cell-free massive MIMO system, it is shown that, compared with the simulated average achievable uplink rate curve, the obtained two approximations (based on (11) with (14),  and (15) with (19), respectively) curves are effective and have the similar growth rate, which only a little gap is exist between the approximation and simulated rate.  In other words, it justifies the effectiveness of {\emph {Theorem 1}} and {\emph {Theorem 2}}.
The similar results can be found for the conventional massive MIMO system, which the upper bound and lower bound curves
are based on averaging out the large-scale fading in the  results (13) and (18), respectively.
Note that the tightness among the upper bound, lower bound, and simulated rate in the conventional massive MIMO system can be analytically proved based on the exact bound expressions (13) and  (18).
It is found that as the number of BS antennas increases, the gap between these two systems is gradually increased for all obtained approximations (bounds), and simulated rate, which shows the cell-free massive MIMO system has a huge development potential for achievable uplink rate improving. Moreover, to explore and analyze the scaling behavior of the achievable uplink rate, two numerical curves which have the asymptotic properties $\Theta\left(\frac{\alpha}{2}\log_2 L\right)$ and $\Theta\left(\log_2 L\right)$, respectively, are plotted.\footnote{Note that to avoid confusion, the detailed parameters of these two curves have not been provided since the provided asymptotic properties are enough. Moreover, we apply this methodology in Fig. 2 (b).} Compared with these two numerical curves, it is shown that, for the cell-free massive MIMO system, the two approximations have an asymptotic lower bound  $\frac{\alpha}{2}\log_2 L$ as $L\rightarrow\infty$ which prove the validity of {\emph {Corollary 1}} and {\emph {Corollary 2}}; for the conventional massive MIMO system, the upper/lower bound and simulated rate do have an asymptotically tight bound $\log_2 L$ which satisfies the conclusion of the discussion after {\emph {Corollary 2}}. In brief, the obtained asymptotic theoretical results are accurate for the rate behavior and can provide key insights to analyze the rate performance.

For the imperfect CSI case of Fig. 2 (b), two approximate curves in cell-free massive MIMO system are based on (30) with (35), and (36) with (43), respectively, while the upper and lower bound curves in conventional massive MIMO system are based on averaging out the large-scale fading
in the results (34) and (42), respectively. It is found that the above mentioned observations in perfect CSI case can also be found in imperfect CSI case, which prove the validity of {\emph {Theorem 3}}, {\emph {Theorem 4}}, {\emph {Corollary 3}}, {\emph {Corollary 4}}, and conclusion of the discussion after {\emph {Corollary 4}}. Only the achievable uplink rate loss is introduced into the imperfect CSI case for both two systems.

\begin{table}[!t]
\tabcolsep 3mm
\caption{List of the average per user RAE (Upper Bound)} \label{tb_variables}
\centering
\begin{tabular}{|c|c|c|c|c|c|c|c|c|c|}
\hline
\multicolumn{2}{|c|}{}& \multicolumn{8}{|c|}{{Average per user RAE}}\\
\hline
\multicolumn{2}{|c|}{{\diagbox{Case}{\!\!\!$L$}}}& 150 & 200 & 250 & 300 & 350 & 400 & 450 & 500\\
\hline
\multirow{2}{*}{\!\!\!\!$\alpha\!\!=\!\!3$\!\!\!\!} & Perfect
                     &1.4580\%   &1.1534\%   &0.9458\%   &0.8076\%   &0.7495\%   &0.6956\%    &0.6560\%   &0.6017\%\\
\cline{2-10}
&\!\!\!\!\!\!Imperfect\!\!\!\!\!\!
                    &1.5568\%   &1.2223\%    &0.9981\%   &0.8498\%   &0.7866\%   &0.7285\%
                    &0.6858\%   &0.6282\%\\
\hline
\multirow{2}{*}{\!\!\!\!$\alpha\!\!=\!\!4$\!\!\!\!} & Perfect
                    &1.4673\%   &1.1680\%    &0.9507\%   &0.8091\%   &0.7509\%   &0.6957\%
                    &0.6535\%   &0.5936\%\\
\cline{2-10}
&\!\!\!\!\!\!Imperfect\!\!\!\!\!\!
                    &1.5184\%   &1.2044\%    &0.9783\%   &0.8315\%   &0.7709\%   &0.7134\%
                    &0.6696\%   &0.6080\%\\
\hline

\end{tabular}
\end{table}

\begin{table}[!t]
\tabcolsep 3mm
\caption{List of the average per user RAE (Lower Bound)} \label{tb_variables}
\centering
\begin{tabular}{|c|c|c|c|c|c|c|c|c|c|}
\hline
\multicolumn{2}{|c|}{}& \multicolumn{8}{|c|}{{Average per user RAE}}\\
\hline
\multicolumn{2}{|c|}{{\diagbox{Case}{\!\!\!$L$}}}& 150 & 200 & 250 & 300 & 350 & 400 & 450 & 500\\
\hline
\multirow{2}{*}{\!\!\!\!$\alpha\!\!=\!\!3$\!\!\!\!} & Perfect
                     &4.9776\%    &4.4547\%   &4.1216\%  &3.9180\%  &3.7434\%  &3.6266\%
                     &3.5628\%    &3.4872\%\\
\cline{2-10}
&\!\!\!\!\!\!Imperfect\!\!\!\!\!\!
                    &5.5632\%     &4.9186\%   &4.5198\%  &4.2742\%  &4.0688\%  &3.9304\%
                    &3.8499\%     &3.7595\%\\
\hline
\multirow{2}{*}{\!\!\!\!$\alpha\!\!=\!\!4$\!\!\!\!} & Perfect
                    &7.0368\%     &6.3729\%   &5.9482\%  &5.7136\%  &5.4854\%  &5.3418\%
                    &5.2495\%     &5.1437\% \\
\cline{2-10}
&\!\!\!\!\!\!Imperfect\!\!\!\!\!\!
                    &7.5423\%     &6.7874\%   &6.3116\%  &6.0459\%  &5.7932\%  &5.6327\%    &5.5269\%     &5.4088\%\\
\hline

\end{tabular}
\end{table}

In the following, to study the influence of the path-loss factor on the obtained approximations, we introduce the relative approximation error (RAE) metric and use it to measure the difference between the simulated bound and the corresponding approximation. The user $k$'s RAE metric is defined as ${\text {RAE}}_{k}\triangleq\mathbb{E}_{\bm \Upsilon}\{|{\rm {TRUE}}_{k}-{\rm {APPRX}}_{k}|/{\rm {TRUE}}_{k}\}$, where ${\rm {TRUE}}_{k}$ denotes any one of the notations $R_{k}^{\text {ub}}$, $\hat{R}_{k}^{\text {ub}}$, $R_{k}^{\text {lb}}$, and $\hat{R}_{k}^{\text {lb}}$. Also, ${\rm {APPRX}}_{k}$ denotes any one of the notations $\widetilde{{R}_{k}^{\text {ub}}}$, $\widetilde{\hat{R}_{k}^{\text {ub}}}$, $\widetilde{R_{k}^{\text {lb}}}$, and $\widetilde{\hat{R}_{k}^{\text {lb}}}$. The average per user RAE values for both perfect CSI and imperfect CSI cases, as well as, different $L$ are presented in Table I (Upper Bound) and Table II (Lower Bound), respectively, where $\rho_u=-10$dB, $\rho_p=0$dB, and $\alpha=3,4$. From Table I, we can see that, the approximation is very accurate based on the average per user RAE performance. Also, the influence of path-loss factor on the RAE performance is negligible. From Table II, we can see that, the gap between the simulated lower bound and the corresponding approximation is negligible except the little gap which exists when $\alpha=4$ and $L=150$. Also, for different path-loss factor, when the number of BS antenna increases, the RAE decreases. In summary, our obtained approximations are accurate and the influence of the path-loss factor on the obtained approximations is negligible.

\begin{figure*}[!t]
\centering
\subfigure[Perfect CSI case] { \label{fig:(a)}
\includegraphics[width=2.8in]{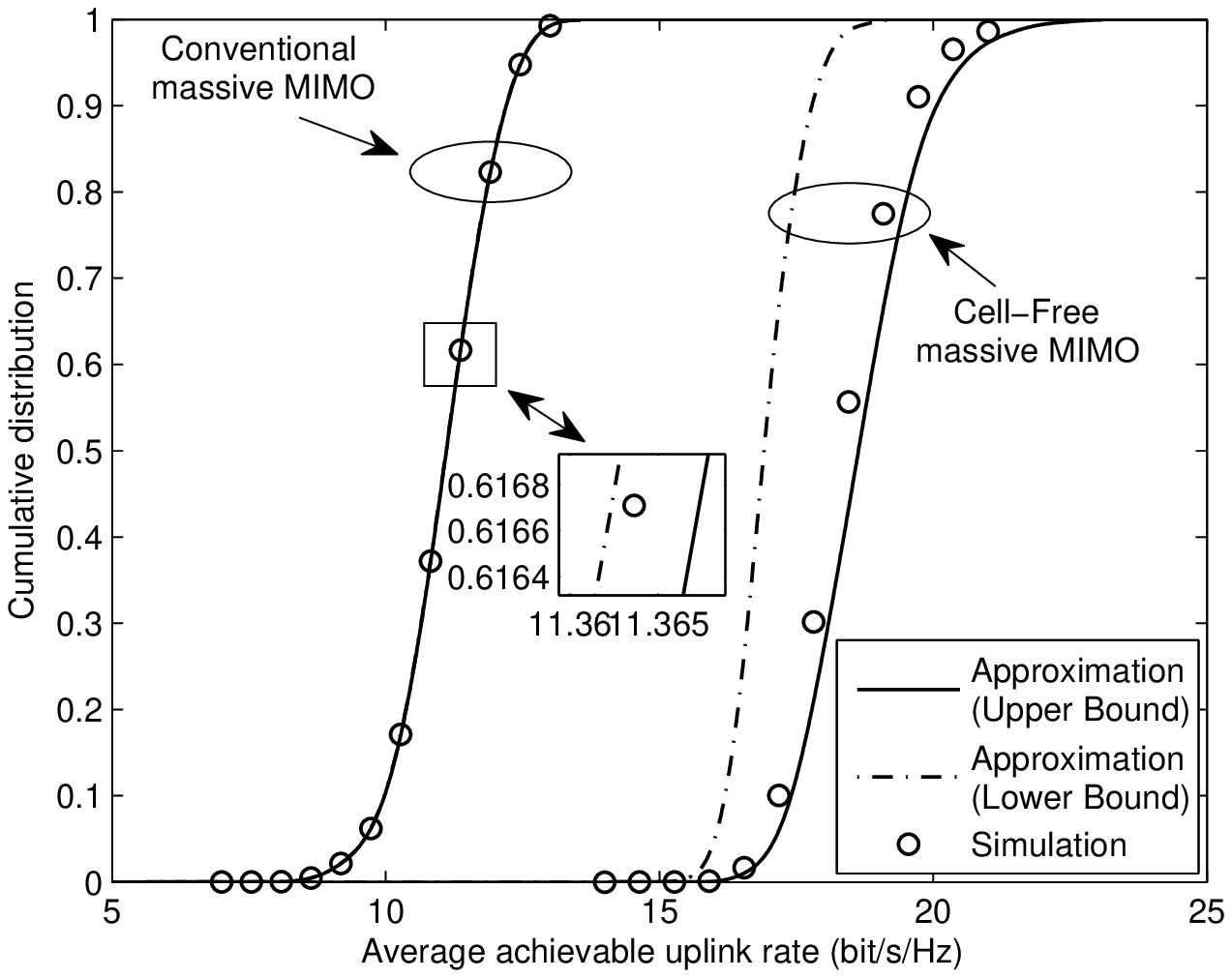}
}
\subfigure[Imperfect CSI case] {\label{fig:(b)}
\includegraphics[width=2.8in]{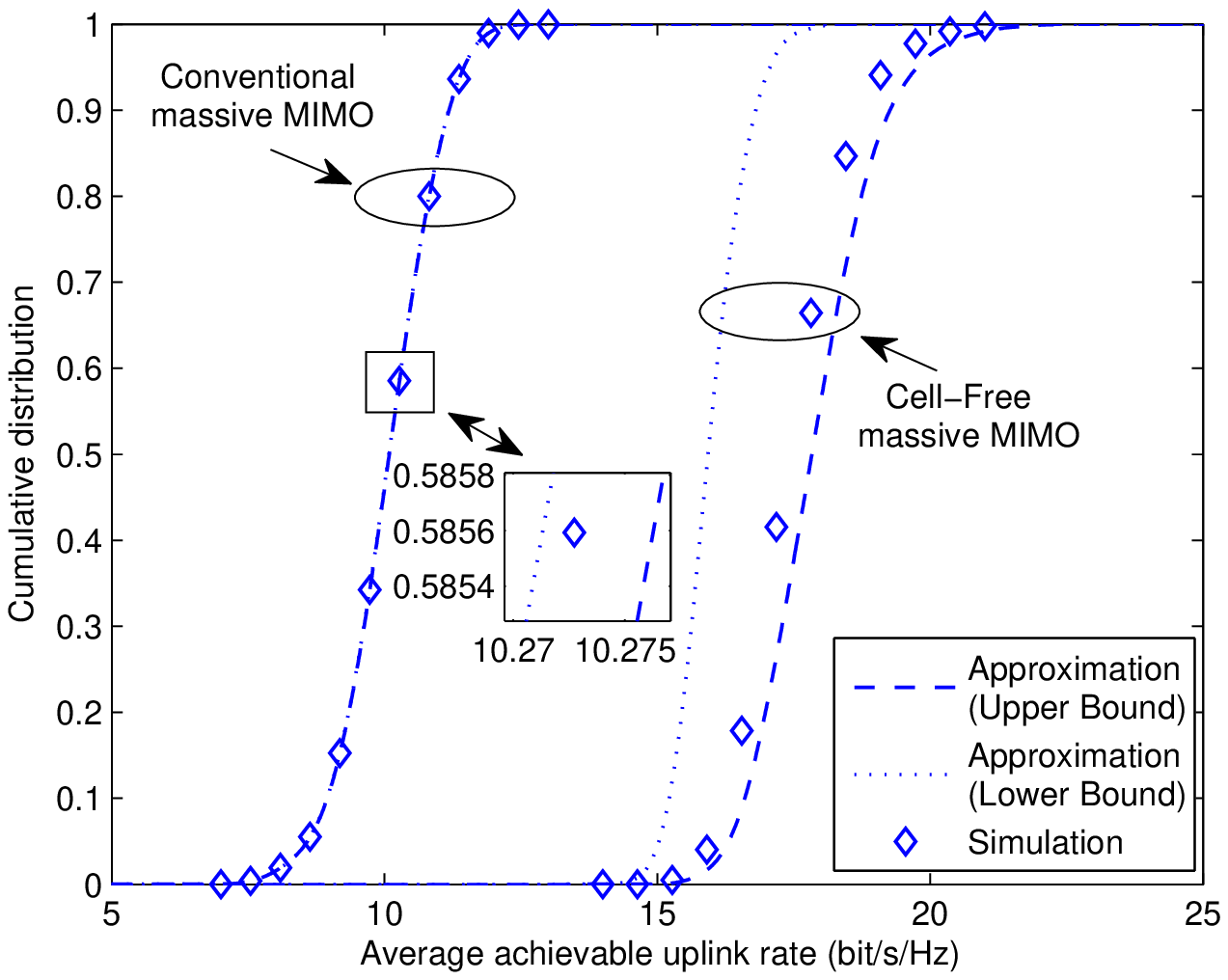}
}
\caption{The CDF of the average achievable uplink rate when $L=300$ and $\rho_u=0$dB for cell-free  and conventional massive MIMO systems with the perfect CSI and imperfect CSI ($\rho_p=10$dB).}
\label{fig2}
\end{figure*}

Fig. 3 presents the cumulative distribution function (CDF) of the average achievable uplink rate with $\rho_u=0$dB and the cell-free massive MIMO system, as well as, the conventional massive MIMO system, for perfect CSI case and imperfect CSI case ($\rho_p=10$dB), under the setting of the number of BS antennas is equal to 300. Moreover, the obtained approximations in cell-free massive MIMO and the bounds in conventional massive MIMO are offered for comparison. Specifically, for the perfect CSI case of Fig. 3 (a), with the cell-free massive MIMO system, the approximate curves line at the either end of the exact simulated achievable uplink rate curve for the whole rate region, and one approximation is tight to the simulation result, which imply that the obtained analytical results in the perfect CSI case  are valid for the whole CDF region. Moreover, compared with the conventional massive MIMO system, the approximate and simulation results in cell-free massive MIMO system show huge advantage in regard of the rate metric. For the imperfect CSI case of Fig. 3 (b), again, the above mentioned observations in perfect CSI case can be deduced that obtained analytical results in the imperfect CSI case are very effective and only achievable uplink rate loss is inevitable due to the channel estimation error.

Fig. 4 investigates the impact of the pilot sequence power $\rho_p$ on the average achievable uplink rate performance.  In this figure, the uplink data power is set to -10dB and the number of BS antennas is set to 300.   For the cell-free massive MIMO system, it shows that the approximations are effective regardless of the value of $\rho_p$ compared with simulation values for both perfect CSI and imperfect CSI cases.  The obtained approximations has the similar growth rate with the exact simulated rate in imperfect CSI case. Moreover, when $\rho_p\rightarrow\infty$, the approximations and simulated rate for imperfect CSI case approach to different constant values,  which match the approximations and simulated rate for perfect CSI case, respectively. In other words, these results showcase the conclusions of (33) and (41) that the performance of imperfect CSI case is always less than the perfect CSI case. The above mentioned results can be found in the conventional massive MIMO system in this figure. Moreover, from this figure, for different $\rho_p$, the performance of cell-free massive MIMO system is far better than the conventional massive MIMO system.

\begin{figure*}[!t]
\makeatletter\def\@captype{figure}\makeatother
\begin{minipage}{0.52\linewidth}
\centering
\includegraphics[width=2.8in]{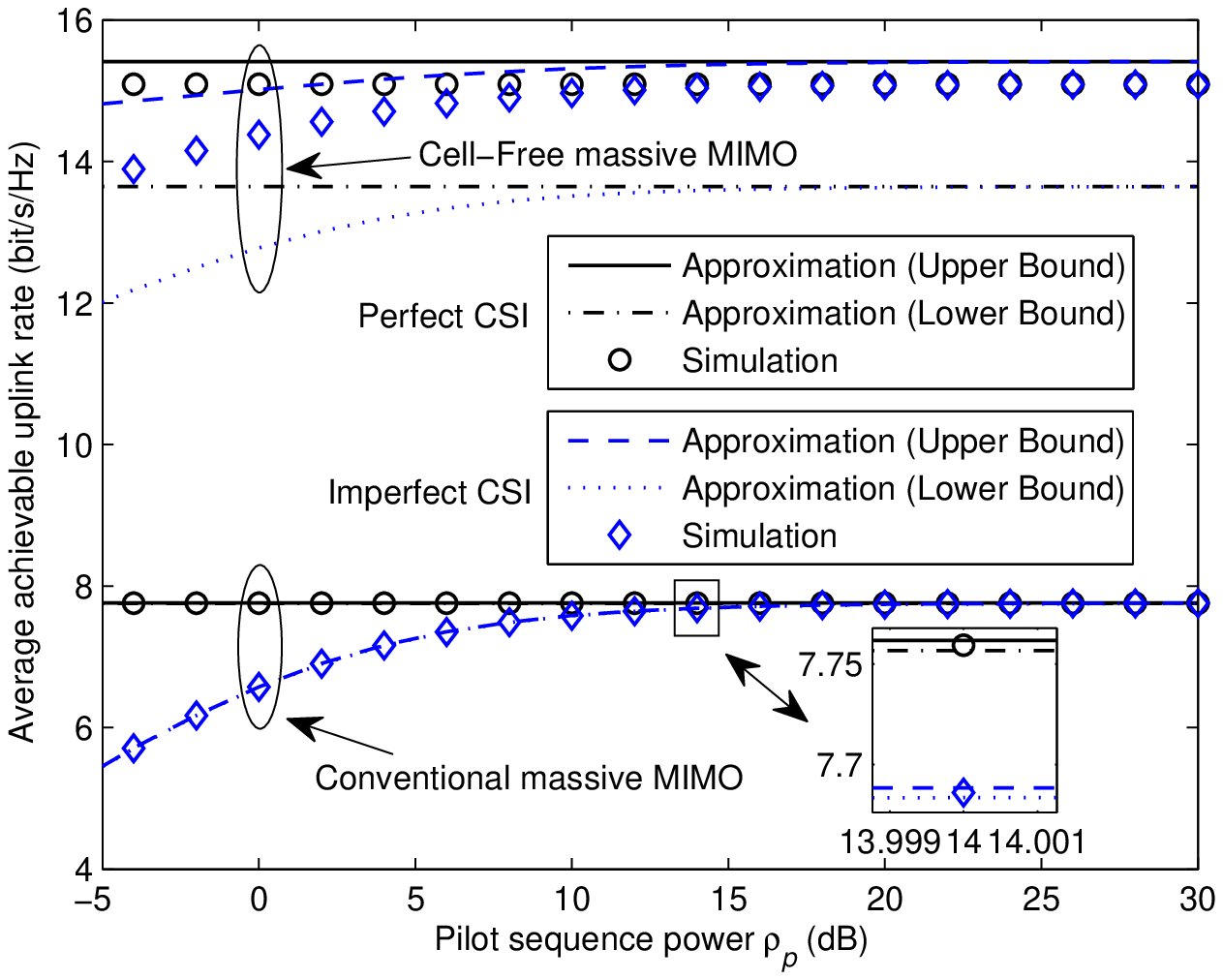}
\caption{The average achievable uplink rate when $\rho_u=-10$dB and $L=300$ for cell-free and conventional massive MIMO systems.}
\label{dropped}
\end{minipage}
\ \
\makeatletter\def\@captype{figure}\makeatother
\begin{minipage}{0.48\linewidth}
\centering
\includegraphics[width=2.5in]{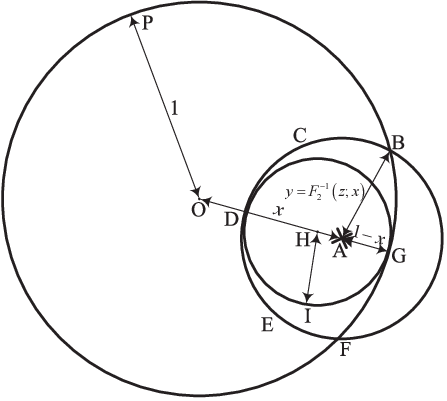}
\caption{Geometric diagram.}
\label{dropped}
\end{minipage}
\end{figure*}

\section{Conclusion}
In this paper, we presented a seminal study for investigating the achievable uplink rate performance of cell-free massive MIMO systems employing ZF detectors. For the perfect CSI case, two novel approximate expressions of the achievable uplink rate were deduced. Moreover, these approximations can converge into the exact bounds in the conventional massive MIMO systems when all BS antennas are co-located. Particularly,
it was found that, with large number of BS antennas $L$, the approximations  by averaging out the large-scale fading had the asymptotic lower bound $\frac{\alpha}{2}\log_2L$, whilst for the upper and lower bounds in conventional massive MIMO systems, it had the asymptotically tight bound $\log_2L$. In other words, these asymptotic results showed the potential of cell-free massive MIMO systems since the path-loss factor $\alpha>2$, except for the free space environment. For the imperfect CSI case, the above mentioned conclusions in perfect CSI case were true of the imperfect CSI case but with a little bit rate loss. In addition, the approximations  in imperfect CSI case converge into the approximations in perfect CSI case when the pilot sequence power $\rho_p$ became infinite.

\appendices

\section{Several Useful Results}
{\emph {Lemma 1 (\cite[Eqs. (57-59)]{Dai11}, \cite[Eqs. (43-46)]{Wang15}, and \cite[Eq. (3.1)]{Yang11}):}} Consider a circular area with unit radius, user $k$ and $L-K+1$ ($L>K-1$) antennas are uniformly distributed within this area. $\rho_k$ denotes the distance between user $k$ and the circular area centre. The PDF of $\rho_k$ is given by
\begin{align}
f_{\rho_k}\left(x\right)\triangleq2x, 0\leq x\leq 1.\tag{46}
\end{align}
Let $d_k^{(l)}(l=1,\ldots,L-K+1)$ is the $l$th minimum access distance from user $k$ to the $L-K+1$ antennas, which its conditional PDF given $\rho_k$ is
\begin{align}
f_{d_k^{(l)}|\rho_k}(y|x)&\!\triangleq\!\frac{(L\!-\!K\!+\!1)!}{(l\!-\!1)!
(L\!-\!K\!+\!1\!-\!l)!}
(F(y;x))^{l-1}\!\left(1\!-\!F(y;x)\right)^{L-K+1-l}\!\!f(y;x), 0\!\leq\! y\!\leq\! 1\!+\!x,\!\tag{47}
\end{align}
where $F(y;x)$ and $f(y;x)$ denote the CDF and PDF of the access distance from user $k$ to any antenna as
\begin{align}
\begin{split}
   F(y;x)&\triangleq\left\{
\begin{array}{*{20}l}
   y^2, &0\leq y\leq 1-x,\\
   \\
   \frac{y^2}{\pi}\arccos\frac{x^2+y^2+1}{2xy}+\frac{1}{\pi}\arccos\frac{1+x^2-y^2}{2x}-\frac{2}{\pi}S_{\Delta}(x,y),
   &1-x<y\leq 1+x,\\
\end{array}\right.
\end{split}\tag{48}
\end{align}
and
\begin{align}
\begin{split}
   f(y;x)&\triangleq\!\left\{
\begin{array}{*{20}l}
   \!\!\!2y, \!&0\leq y\leq 1-x,\\
   \\
   \!\!\!\frac{2y}{\pi}\arccos\frac{x^2+y^2+1}{2xy}, \!&1-x<y\leq 1+x,\\
\end{array}\right.
\end{split}\tag{49}
\end{align}
respectively. Besides, $S_{\Delta}(x,y)$ is defined as
\begin{align}
S_{\Delta}(x,y)&\triangleq\sqrt{\frac{x+y+1}{2}\left(\frac{x+y+1}{2}-1\right)
\left(\frac{x+y+1}{2}-x\right)
\left(\frac{x+y+1}{2}-y\right)}.\tag{50}
\end{align}

{\emph {Lemma 2 (\cite[Lemma 2]{Li18} and \cite[Lemma 6 \& Proposition 8]{Heath11}):}}
If $X_{l}$ ($l=1,\ldots, L$) is an independent Gamma distributed random variable with a shape parameter $\eta_{l}$ and a scale parameter $\theta_{l}$, i.e., $X_{l}\sim\Gamma(\eta_{l}, \theta_{l})$, then, the sum random variable $X=\sum\limits_{l=1}^{L}X_{l}$  can be approximated by a Gamma distributed random variable $\hat{X}\sim\Gamma(\hat{\eta}, \hat{\theta})$, which  $\hat{\eta}$ and $\hat{\theta}$ are defined as $\hat{\eta}\triangleq{\left(\sum\limits_{l=1}^{L}\eta_{l}\theta_{l}\right)^2}/{\sum\limits_{l=1}^{L}\eta_{l}\theta_{l}^{2}}$
and $\hat{\theta}\triangleq
{\sum\limits_{l=1}^{L}\eta_{l}\theta_{l}^{2}}/
{\sum\limits_{l=1}^{L}\eta_{l}\theta_{l}}$, respectively.

{\emph {Proposition 1:}} Consider the same conditions as in {\emph {Lemma 1}}, for fixed $l$, as $L\rightarrow\infty$,
the expectation of $\left(d_k^{(l)}\right)^{\alpha}$ has the following asymptotic property\footnote{Note that the specifical case $l=1$ has been provided in \cite{Dai14} with brief proof.
Here, we  provide detailed result for general case which is very generic.}
\begin{align}
Q_l(L)\triangleq\mathbb{E}_{d_k^{(l)}}\left\{\left(d_k^{(l)}\right)^{\alpha}\right\}
=\Theta\left(L^{-\frac{\alpha}{2}}\right),\tag{51}
\end{align}
where $\alpha$ is the path-loss factor which satisfies $\alpha>2$, except for
the free space environment.

{\emph {Proof:}} Based on the definition of the expectation, we start by writing
\begin{align}
Q_l(L)
&=\int_{0}^{1}\!\!\!
\int_{0}^{1+x}\!\!\!\!\!\!\!\!\!f_{\rho_k}(x)f_{d_k^{(l)}|\rho_k}(y|x)y^{\alpha}
\mathrm{d}x\mathrm{d}y.\tag{52}
\end{align}
To evaluate the integral (52), by substituting (47) into $Q_l(L)$ and defining $z\triangleq F(y;x)$, after some algebraic manipulation, we have
\begin{align}
Q_l(L)
&=\frac{(L-K+1)!}{(l-1)!(L-K+1-l)!}
\int_{0}^{1}\!\!\!\!
\int_{0}^{1}\!\!\!f_{\rho_k}(x)z^{l-1}\left(1-z\right)^{L-K+1-l}
\left(F^{-1}(z;x)\right)^{\alpha}\!\mathrm{d}x\mathrm{d}z
\notag\\
&=\frac{(L-K+1)!}{(l-1)!(L-K+1-l)!}
\int_{0}^{1}\!\!\!\!f_{\rho_k}(x)
\Bigg(\overbrace{\int_{0}^{1}\!\!\!
z^{l-1}\left(1-z\right)^{L-K+1-l}
\left(F^{-1}(z;x)\right)^{\alpha}\!\mathrm{d}z}^{Q_l(L;x)}\Bigg)\mathrm{d}x
,\tag{53}
\end{align}
where $F^{-1}(\cdot)$ denotes the inverse function of $F(\cdot)$.

To analyze the scaling behavior of $Q_l(L)$, we firstly investigate the term $Q_l(L;x)$.
Based on the structure of $F(y;x)$ in (48), it is known that $F(0;x)=0$, $F(1-x;x)=(1-x)^2$, and $F(1+x;x)=1$. Then, for convenience, when $y\in[0,1-x]$, we denote $F(y;x)$ as $F_1(y;x)\triangleq y^2\in[0,(1-x)^2]$. Hence, when $z\in[0,(1-x)^2]$, $F_1^{-1}(z;x)\in[0,1-x]$. When $y\in(1-x,1+x]$, we denote $F(y;x)$ as $F_2(y;x)\triangleq\frac{y^2}{\pi}\arccos\frac{x^2+y^2+1}{2xy}+\frac{1}{\pi}\arccos\frac{1+x^2-y^2}{2x}-\frac{2}{\pi}S_{\Delta}(x,y)
\in((1-x)^2,1]$. Hence, when $z\in((1-x)^2,1]$, $F_2^{-1}(z;x)\in(1-x,1+x]$. Next, for the term $Q_l(L;x)$, after much algebraic manipulation, we perform the following sequence of operations as
\begin{align}
Q_l(L;x)&=\!\!\!\underbrace{\int_{0}^{1}\!\!\!
z^{l-1}\left(1-z\right)^{L-K+1-l}
\left(F_1^{-1}(z;x)\right)^{\alpha}\mathrm{d}z}_{Q_{l1}(L;x)}\notag\\
&\ \ \ +\underbrace{\int_{(1-x)^2}^{1}\!\!\!\!\!\!\!\!\!
z^{l-1}\left(1-z\right)^{L-K+1-l} \left(\left(F_2^{-1}(z;x)\right)^{\alpha}-\left(F_1^{-1}(z;x)\right)^{\alpha}\right)
\mathrm{d}z}_{Q_{l2}(L;x)}.\tag{54}
\end{align}

In the following, to investigate the relationship between $Q_{l1}(L;x)$ and $Q_{l2}(L;x)$, we firstly consider the case $0<x\leq1$ and $z\in((1-x)^2,1]$.  Consider a circle area with center $\text{O}$ and radius $\overline{{\text {OP}}}=1$, which is shown in Fig. 5. A user is placed at the point A with $\overline{{\text {OA}}}=x\ (0<x\leq1)$. Also, the extension line of $\overline{{\text {OA}}}$ meets the circle O at the point G, i.e, $\overline{{\text {AG}}}=1-x$. Then, since the BS antennas are uniformly distributed, given the user located at the point A, it is obvious that $z=F_2(y;x)={S_{\text{BCDEFG}}}/{\pi}$, where $S_{\text{BCDEFG}}$ is shown in Fig. 5, which is the intersection area of circle O and the circle with center A and radius $\overline{{\text {AB}}}=\overline{{\text {AD}}}=y=F_2^{-1}(z;x)$. On one hand, it can be found that the area of circle  A, $S_{{\text A}}$, is great than $S_{\text{BCDEFG}}$, clearly, $S_{{\text A}}=\pi y^2=\pi (F_2^{-1}(z;x))^2>\pi F_2(y;x)=\pi z$. On the other hand, we can also draw a circle with the center H ($\overline{{\text {DG}}}=\overline{{\text {DA}}}+\overline{{\text {AG}}}=2\overline{{\text {HI}}}=2\overline{{\text {HD}}}=2\overline{{\text {HG}}}={F_2^{-1}(z;x)+(1-x)}$). It can be shown that the circle H meets circle A and circle O only at the points D and G, respectively. Hence, $S_{\text{BCDEFG}}=\pi z\geq S_{{\text H}}=\pi \left(({F_2^{-1}(z;x)+(1-x)})/2\right)^2$, where the ``$=$" in "$\geq$" is reached when $z=1$.
Finally, based on above mentioned, we obtain
\begin{align}
\pi\left(\frac{F_2^{-1}(z;x)+(1-x)}{2}\right)^2
\leq\pi z<\pi\left(F_2^{-1}(z;x)\right)^2,\tag{55}
\end{align}
where the ``$=$" in "$\leq$" is reached when $z=1$. Then (55) yields
\begin{align}
1-x&<F_1^{-1}(z;x)<
F_2^{-1}(z;x)\leq2F_1^{-1}(z;x)-(1-x).\tag{56}
\end{align}

Hence,  with the aid of (56), when $0<x\leq1$, the term $Q_{l2}(L;x)$ is upper bounded by
\begin{align}
Q_{l2}(L;x)&<\left(2^\alpha-1\right)\int_{(1-x)^2}^{1}
z^{l-1}\!\left(1\!-\!z\right)^{L-K+1-l}\! \left(F_1^{-1}(z;x)\right)^{\alpha}\!
\mathrm{d}z\notag\\
&<\left(2^\alpha-1\right)Q_{l1}(L;x).\tag{57}
\end{align}
Moreover, when $x=0$, $Q_{l2}(L;0)=0$ and $Q_{l1}(L;0)>0$. Hence, it is proved that
\begin{align}
Q_{l2}(L;x)=O\left(Q_{l1}(L;x)\right), \forall x\in[0,1].\tag{58}
\end{align}
Next, based on (48), the integral $Q_{l1}(L;x)$ is now evaluated as
\begin{align}
Q_{l1}(L;x)&=\int_{0}^{1}\!\!\!
z^{l+\frac{\alpha}{2}-1}\left(1-z\right)^{L-K+1-l}\mathrm{d}z\notag\\
&={\rm B}\left(l+\frac{\alpha}{2}, L-K+1-(l-1)\right),\tag{59}
\end{align}
where ${\rm B}\left(x, y\right)\triangleq\int_{0}^{1}t^{x-1}(1-t)^{y-1}\mathrm{d}t$ is the Beta function.
Note that when $y$ is large and $x$ is fixed, ${\rm B}\left(x,y\right)\rightarrow\Gamma(x)y^{-x}$, hence when $L\rightarrow\infty$, we have
\begin{align}
Q_{l1}(L;x)&\rightarrow\notag\frac{\Gamma(l+\frac{\alpha}{2})}{(L-K+1-(l-1))^{l+\frac{\alpha}{2}}}\notag\\
&=\Theta\left(L^{-(l+\frac{\alpha}{2})}\right).\tag{60}
\end{align}
Then, based on (58) and (60), we have
\begin{align}
Q_{l}(L;x)&=\Theta\left(L^{-(l+\frac{\alpha}{2})}\right).\tag{61}
\end{align}
Finally, the conclusion (51) is obtained by substituting (61) into (53), unitizing $\frac{(L-K+1)!}{(l-1)!(L-K+1-l)!}=\Theta\left(L^{l}\right)$, and simplifying.
\hfill\rule{3mm}{3mm}

\section{Proof of Theorem 1}
Based on the Jensen's inequality \cite{Ngo13}, (7) is upper bounded by
\begin{align}
R_{k}&\leq
\log_2\left(1+\rho_{u}
\mathbb{E}_{\bf H}\left\{\frac{1}{\|{\bf a}_{k}\|_2^2}\right\}
\right)\triangleq R_{k}^{\text {ub}},\tag{62}
\end{align}
where the ``$=$" in ``$\leq$" is reached when $\frac{1}{\|{\bf a}_{k}\|_2^2}$ is a constant. Hence, the key issue is to obtain the distribution of $\frac{1}{\|{\bf a}_{k}\|_2^2}$.
By utilizing the methodology of \cite[Appendix D]{Wang15}, when $L$ is large, we have the following approximation \begin{align}
\frac{1}{\|{\bf a}_{k}\|_2^2}&\approx\|{\bf g}_{k}\|_2^2
-\sum\limits_{n\neq k}^{K}|g_{l_{n}^{\star},k}|^2.
\tag{63}
\end{align}
Also, when $L\gg K$, we note that the elements of the set $\{l_{n}^{\star}|\forall n\neq k\}$ are almost reciprocal. Hence, we can approximate the set $\{l_{n}^{\star}|\forall n\neq k\}$ as the set ${\mathcal {A}}_{k}$, where ${\mathcal {A}}_{k}$ has been defined as in (12). In other words, we obtain
\begin{align}
\sum\limits_{n\neq k}^{K}|g_{l_{n}^{\star},k}|^2\thickapprox
\sum\limits_{\tilde{l}\in{\mathcal {A}}_{k}}|g_{\tilde{l},k}|^2.
\tag{64}
\end{align}
Finally, (11) is obtained by  substituting (2), (63), and (64) into (62) and simplifying.
\hfill\rule{3mm}{3mm}

\section{Proof of Corollary 1}
Based on {\emph {Theorem 1}}, it is easy to obtain the approximation (14). In the following, the asymptotic behavior of $\widetilde{\bar{R}_{k}^{\text {ub}}}$ is proved.

First, by denoting $\gamma_{k}^{(l)}$ and $d_{k}^{(l)}$ as the $l$th maximum large-scale fading coefficient and the $l$th minimum access distance from user $k$ to the antennas belonged to the set ${\mathcal {L}}_{k}$, respectively,
$\widetilde{\bar{R}_{k}^{\text {ub}}}$ is rewritten and lower bounded by
\begin{align}
\widetilde{\bar{R}_{k}^{\text {ub}}}&=\mathbb{E}_{\bm \Upsilon}
\left\{\log_2\left(1+\rho_{u}\sum\limits_{l=1}^{|{\mathcal {L}}_{k}|}\gamma_{k}^{(l)}\right)\right\}\notag\\
&>\mathbb{E}_{\bm \Upsilon}
\left\{\log_2\left(1+\rho_{u}\gamma_{k}^{(1)}\right)\right\}\triangleq\widetilde{\bar{R}_{k}^{\text {ub-lb}}}
.\tag{65}
\end{align}

Then, when $L$ is large, $|{\mathcal {L}}_{k}|\rightarrow L-K+1$ and ${\mathcal {L}}_{k}$ can be regarded and composed of $L-K+1$ uniformly distributed BS antennas in the circular area with unit radius \cite{Wang15}. Hence, with the aid of (3) and  the Jensen's inequality, $\widetilde{\bar{R}_{k}^{\text {ub-lb}}}$ is rewritten and lower bounded by
\begin{align}
\widetilde{\bar{R}_{k}^{\text {ub-lb}}}&=\int_{0}^{1}\!\!\!
\int_{0}^{1+x}\!\!\!f_{\rho_k}(x)f_{d_k^{(1)}|\rho_k}(y|x)\log_2\left(1+\rho_uy^{-\alpha}\right)
\mathrm{d}x\mathrm{d}y\notag\\
&=\mathbb{E}_{d_k^{(1)}}\left\{\log_2\left(1+\frac{\rho_u}{\left(d_k^{(1)}\right)^\alpha}\right)\right\}
\geq\log_2\left(1+\frac{\rho_u}{Q_1(L)}\right),
\tag{66}
\end{align}
where the ``$=$" in "$\geq$" is reached when $d_k^{(1)}$ is a constant.
The expressions of $f_{\rho_k}(x)$ and $f_{d_k^{(1)}|\rho_k}(y|x)$ can be found in (46) and (47), respectively, in the {\emph {Lemma 1}} of Appendix A. Besides, $Q_1(L)$ is the special case of (51) in the {\emph {Proposition 1}} of Appendix A, which has the property $Q_1(L)=\Theta\left(L^{-\frac{\alpha}{2}}\right)$. In other words, $Q_1(L)$ has an asymptotically tight bound $L^{-\frac{\alpha}{2}}$.
Finally, this asymptotic conclusion is obtained based on the relationship among the $Q_1(L)$, (65), and (66). \hfill\rule{3mm}{3mm}

\section{Proof of Theorem 2}
Based on the Jensen's inequality, (7) is lower bounded by
\begin{align}
R_{k}&\geq
\log_2\left(1+
\frac{\rho_{u}}{\mathbb{E}_{\bf H}\left\{\|{\bf a}_{k}\|_2^2\right\}}
\right)\triangleq R_{k}^{\text {lb}},\tag{67}
\end{align}
where the ``$=$" in ``$\geq$" is reached when ${\|{\bf a}_{k}\|_2^2}$ is a constant.
Hence, the key issue is to obtain the distribution of $\|{\bf a}_{k}\|_2^2$. By substituting (63) and (64) into it, $\|{\bf a}_{k}\|_2^2$ is approximated by
\begin{align}
{\|{\bf a}_{k}\|_2^2}\approx
\frac{1}
{\sum\limits_{\tilde{l}\in{\mathcal {L}}_{k}}|g_{\tilde{l},k}|^2}.\tag{68}
\end{align}
Based on (2), it is note that $\forall \tilde{l}\in{\mathcal {L}}_{k}$,
$|g_{\tilde{l},k}|^2\sim\Gamma(1, \gamma_{\tilde{l},k})$ where the Gamma distributed
random variable $x\sim\Gamma(\eta, \theta)$ has the PDF $f(x)=\frac{x^{\eta-1}e^{\frac{x}{\theta}}}{\theta^{\eta}\Gamma(\eta)}, x\geq 0$.
Then, via the application of {\emph {Lemma 2}} in Appendix A, $\sum\limits_{\tilde{l}\in{\mathcal {L}}_{k}}|g_{\tilde{l},k}|^2$ can be approximated by the Gamma distributed
random variable $\Lambda_k\sim\Gamma(\Psi_{k}, \Phi_{k})$, where $\Psi_{k}$ and $\Phi_{k}$ have defined as in (16) and (17), respectively. Hence, based on the above mentioned results, after some algebraic manipulation, we perform the following sequence of operations
\begin{align}
\mathbb{E}_{\bf H}\left\{{\|{\bf a}_{k}\|_2^2}\right\}
&\approx\mathbb{E}_{\Lambda_k}\left\{\frac{1}{\Lambda_k}\right\}
=\frac{1}{\Phi_{k}\left(\Psi_{k}-1\right)}.\tag{69}
\end{align}
Finally, (15) is obtained by  substituting (69) into (67) and simplifying.   \hfill\rule{3mm}{3mm}

\section{Proof of Corollary 2}
By following the similar methodology of {\emph { Corollary 1}}, the asymptotic behavior of $\widetilde{\bar{R}_{k}^{\text {lb}}}$ is provided based on the subsequent steps.

First,  by adopting the similar sorting method of large-scale fading coefficient as in {\emph {Corollary 1}}, $\widetilde{{R}_{k}^{\text {lb}}}$ is rewritten as
\begin{align}
\widetilde{{R}_{k}^{\text {lb}}}&=\mathscr{F}\left(\gamma_{k}^{(1)},\gamma_{k}^{(2)},\ldots,\gamma_{k}^{(|{\mathcal {L}}_{k}|-1)},\gamma_{k}^{(|{\mathcal {L}}_{k}|)}\right)\notag\\
&=\log_2\left(1+\rho_{u}\Phi_{k}^{\text {sort}}\left(\Psi_{k}^{\text {sort}}-1\right)\right),\tag{70}
\end{align}
where $\mathscr{F}(x_1,x_2,\ldots,x_{K-1},x_{K})
\triangleq\log_2(1+\rho_{u}({\sum\limits_{k=1}^{K}x_{k}^2}/
                            {\sum\limits_{k=1}^{K}x_{k}})
                            ({(\sum\limits_{k=1}^{K}x_{k})^2}/
                                       {\sum\limits_{k=1}^{K}x_{k}^2}
                                        -1))$ and
\begin{align}
&\Phi_{k}^{\text {sort}}\triangleq
\frac{\sum\limits_{l=1}^{|{\mathcal {L}}_{k}|}\left(\gamma_{k}^{(l)}\right)^2}
{\sum\limits_{l=1}^{|{\mathcal {L}}_{k}|}\gamma_{k}^{(l)}},\tag{71}\\
&\Psi_{k}^{\text {sort}}\triangleq
\frac{\left(\sum\limits_{l=1}^{|{\mathcal {L}}_{k}|}\gamma_{k}^{(l)}\right)^2}
{\sum\limits_{l=1}^{|{\mathcal {L}}_{k}|}\left(\gamma_{k}^{(l)}\right)^2}.\tag{72}
\end{align}
Then, $\forall t=1,\ldots,|{\mathcal {L}}_{k}|$, after lengthy algebraic manipulations, we have
\begin{align}
\frac{\partial{\widetilde{{R}_{k}^{\text {lb}}}}}
{\partial{\gamma_{k}^{(t)}}}
&=\frac{\rho_u\left(\sum\limits_{l\neq t}^{|{\mathcal {L}}_{k}|}\left(\gamma_{k}^{(l)}\right)^2
+\left(\sum\limits_{l\neq t}^{|{\mathcal {L}}_{k}|}\gamma_{k}^{(l)}\right)^2\right)}
{\left(1+\rho_{u}\Phi_{k}^{\text {sort}}\left(\Psi_{k}^{\text {sort}}-1\right)\right)\left(\sum\limits_{l=1}^{|{\mathcal {L}}_{k}|}\gamma_{k}^{(l)}\right)^{2}}
\log_2 e>0.\tag{73}
\end{align}
By substituting (70) into $\widetilde{\bar{R}_{k}^{\text {lb}}}$ and unitizing (73), we get
\begin{align}
\widetilde{\bar{R}_{k}^{\text {lb}}}
&>\mathbb{E}_{\bm \Upsilon}
\left\{\mathscr{F}\left(\gamma_{k}^{(2)},\gamma_{k}^{(2)},\ldots,0,0\right)\right\}\notag\\
&=\mathbb{E}_{\bm \Upsilon}
\left\{\log_2\left(1+\rho_{u}\gamma_{k}^{(2)}\right)\right\}\triangleq\widetilde{\bar{R}_{k}^{\text {lb-lb}}}.\tag{74}
\end{align}

Next, utilizing (3) and the methodology of (66),  after some manipulations, $\widetilde{\bar{R}_{k}^{\text {lb-lb}}}$ is lower bounded by
\begin{align}
\widetilde{\bar{R}_{k}^{\text {lb-lb}}}&=\int_{0}^{1}\!\!\!
\int_{0}^{1+x}\!\!\!\!\!\!\!\!\!f_{\rho_k}(x)f_{d_k^{(2)}|\rho_k}(y|x)\log_2\left(1+\rho_uy^{-\alpha}\right) \mathrm{d}x\mathrm{d}y
\geq\log_2\left(1+\frac{\rho_u}{Q_2(L)}\right),\tag{75}
\end{align}
where the ``$=$" in "$\geq$" is reached when $d_k^{(2)}$ is a constant, $f_{d_k^{(2)}|\rho_k}(y|x)$ is the conditional PDF of $d_k^{(2)}$ given $\rho_{k}$ which is the special case of (47) in the {\emph {Lemma 1}} of Appendix A, and $Q_2(L)=\Theta\left(L^{-\frac{\alpha}{2}}\right)$ is the special case of (51) in the {\emph {Proposition 1}} of Appendix A.
Finally,  the asymptotic performance of {\emph {Corollary 2}} can be obtained by the above mentioned intermediate steps. \hfill\rule{3mm}{3mm}

%
\section*{Acknowledgment}
The authors would like to thank Prof. Lin Dai, from City University of Hong Kong, for her generous help in improving this paper.


\ifCLASSOPTIONcaptionsoff
  \newpage
\fi

\footnotesize
\bibliographystyle{ieee}
\bibliography{CASSreference}

\end{document}